\documentstyle[epsf,12pt]{article}
\begin{document}
\begin{center}
\Large{\bf A Hole torn by QCD Fields.}\\
\large{R.S. Longacre$^a$\\
$^a$Brookhaven National Laboratory, Upton, NY 11973, USA}
\end{center}
 
\begin{abstract}
In this paper we determine that a local negative correlation(the Hole) in 
$\Delta \eta$ $\Delta \phi$ space of Au + Au collisions at $\sqrt{s_{NN}}$=200 
GeV is caused by strong color QCD electric and magnetic fields which are 
present in the Glasma Flux Tubes that are generated by the initial conditions. 
\end{abstract}
 
\section{Introduction and review of the model} 

In this paper we discuss a Glasma Flux Tube Model(GFTM)\cite{Dumitru}
which has tubes of parallel strong color QCD electric and magnetic fields that 
are generated by the initial conditions of Au + Au central collisions at 
$\sqrt{s_{NN}}$=200 GeV.

The paper is organized in the following manner:

Sec. 1 is the introduction and review of the model. Sec. 2 discuss two particle
angular correlation in the model. Sec. 3 gives the properties of the
GFTM. Sec. 4 is a comparison of charge independent two particle angular 
correlation between the model and the Au + Au data. Sec. 5 is a comparison of
the charge dependent correlation of the signal coming from a single flux tube
with the Au + Au data. Sec. 6 is about QCD and the strong CP violation term. 
Sec. 7 considers color electric fields and data. Sec. 8 considers color 
magnetic fields and data. Sec. 9 presents the summary and discussion.

\subsection{Glasma Flux Tube Model} 

A glasma flux tube model (GFTM)\cite{Dumitru} that had been developed considers
that the wave functions of the incoming projectiles, form sheets of color glass 
condensates (CGC)\cite{CGC} that at high energies collide, interact, and 
evolve into high intensity color electric and magnetic fields. This collection 
of primordial fields is the Glasma\cite{Lappi,Gelis}, and initially it is 
composed of only rapidity independent longitudinal color electric and magnetic 
fields. An essential feature of the Glasma is that the fields are localized 
in the transverse space of the collision zone with a size of 1/$Q_s$. $Q_s$ is
the saturation momentum of partons in the nuclear wave function. These
longitudinal color electric and magnetic fields generate topological 
Chern-Simons charge\cite{Simons} which becomes a source for particle 
production.

The transverse space is filled with flux tubes of large longitudinal extent 
but small transverse size $\sim$$Q^{-1}_s$. Particle production from a flux 
tube is a Poisson process, since the flux tube is a coherent state. 
The flux tubes at the center of the transverse plane interact with each other
through plasma instabilities\cite{Lappi,Romatschke1} and create a locally 
thermalized system, where partons emitted from these flux tubes are locally
equilibrated. A hydro system with transverse flow builds causing a radially
flowing blast wave\cite{Gavin}. The flux tubes that are near the surface of the
fireball get the largest radial flow and are emitted from the surface.
 
$Q_s$ is around 1 GeV/c thus the transverse size of the flux tube is about 
1/4 fm. The flux tubes near the surface are initially at a radius $\sim$5 fm. 
The $\phi$ angle wedge of the flux tube is $\sim$1/20 radians or 
$\sim$$3^\circ$. Thus the flux tube initially has a narrow range in $\phi$. The
width in $\Delta \eta$ correlation of particles results from the 
independent longitudinal color electric and magnetic fields that created the 
Glasma flux tubes. In this paper we relate particle production from the 
surface flux tube to a related model Parton Bubble Model(PBM)\cite{PBM}. It was
shown in Ref.\cite{PBMtoGFTM} that for central Au + Au collisions at 
$\sqrt{s_{NN}}$=200 the PBM is a good approximation to the GFTM surface flux 
tube formation. 

The flux tubes on the surface turn out to be on the average 12 in number. They
form an approximate ring about the center of the collision see Figure 1. The 
twelve tube ring creates the average behavior of tube survival near the 
surface of the expanding fire ball of the blast wave. The final state surface 
tubes that emit the final state particles at kinetic freezeout are given
by the PBM. One should note that the blast wave surface is moving at its 
maximum velocity at freezeout (3c/4). 

\begin{figure}
\begin{center}
\mbox{
   \epsfysize 5.0in
   \epsfbox{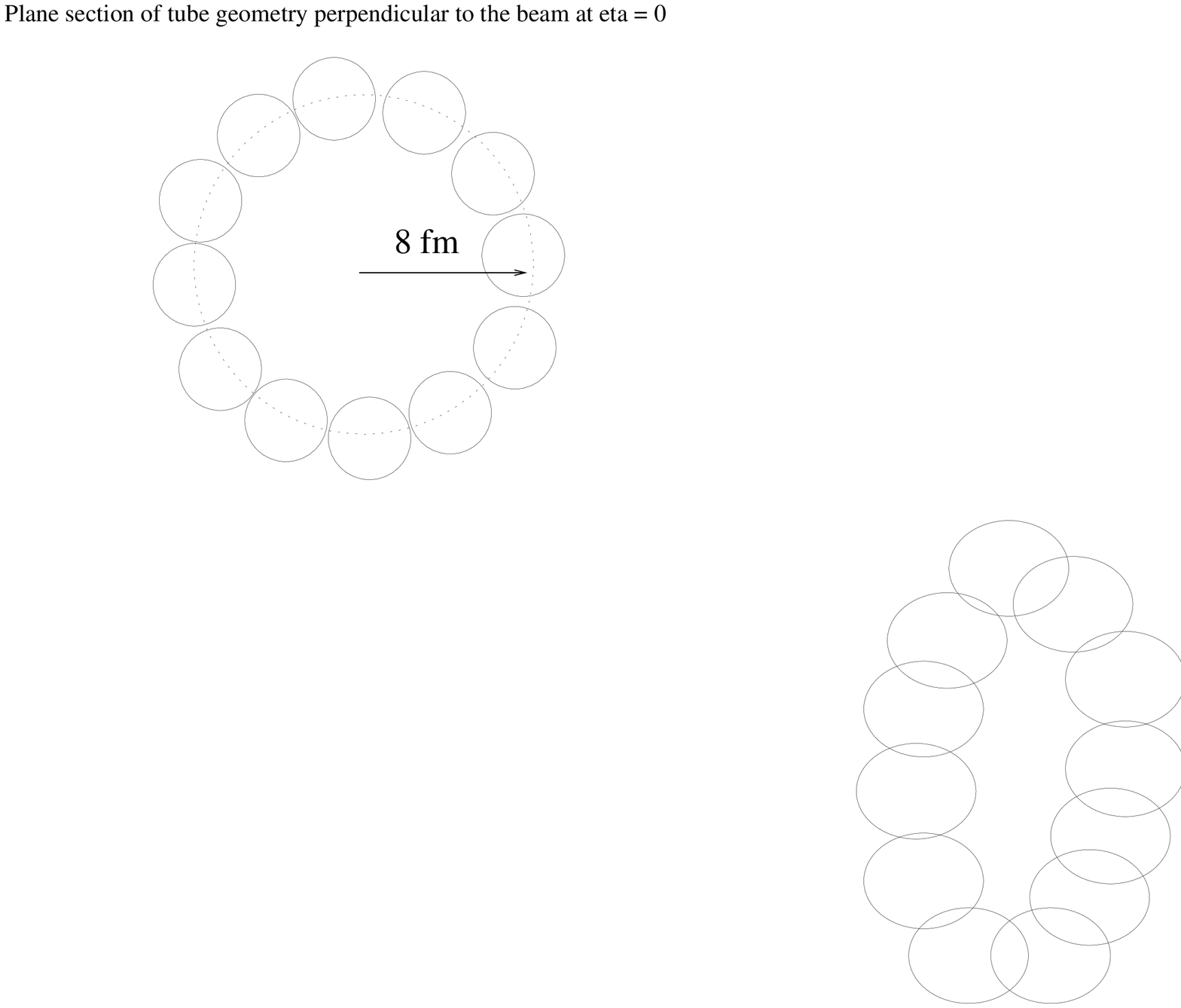}}
\end{center}
\vspace{2pt}
\caption{The tube geometry is an 8 fm radius ring perpendicular to and 
centered on the beam axis. It is composed of twelve adjacent 2 fm radius 
circular tubes elongated along the beam direction as part of the flux
tube geometry. We project on a plane section perpendicular to the beam axis.}
\label{fig1}
\end{figure}

\begin{figure}
\begin{center}
\mbox{
   \epsfysize 5.0in
   \epsfbox{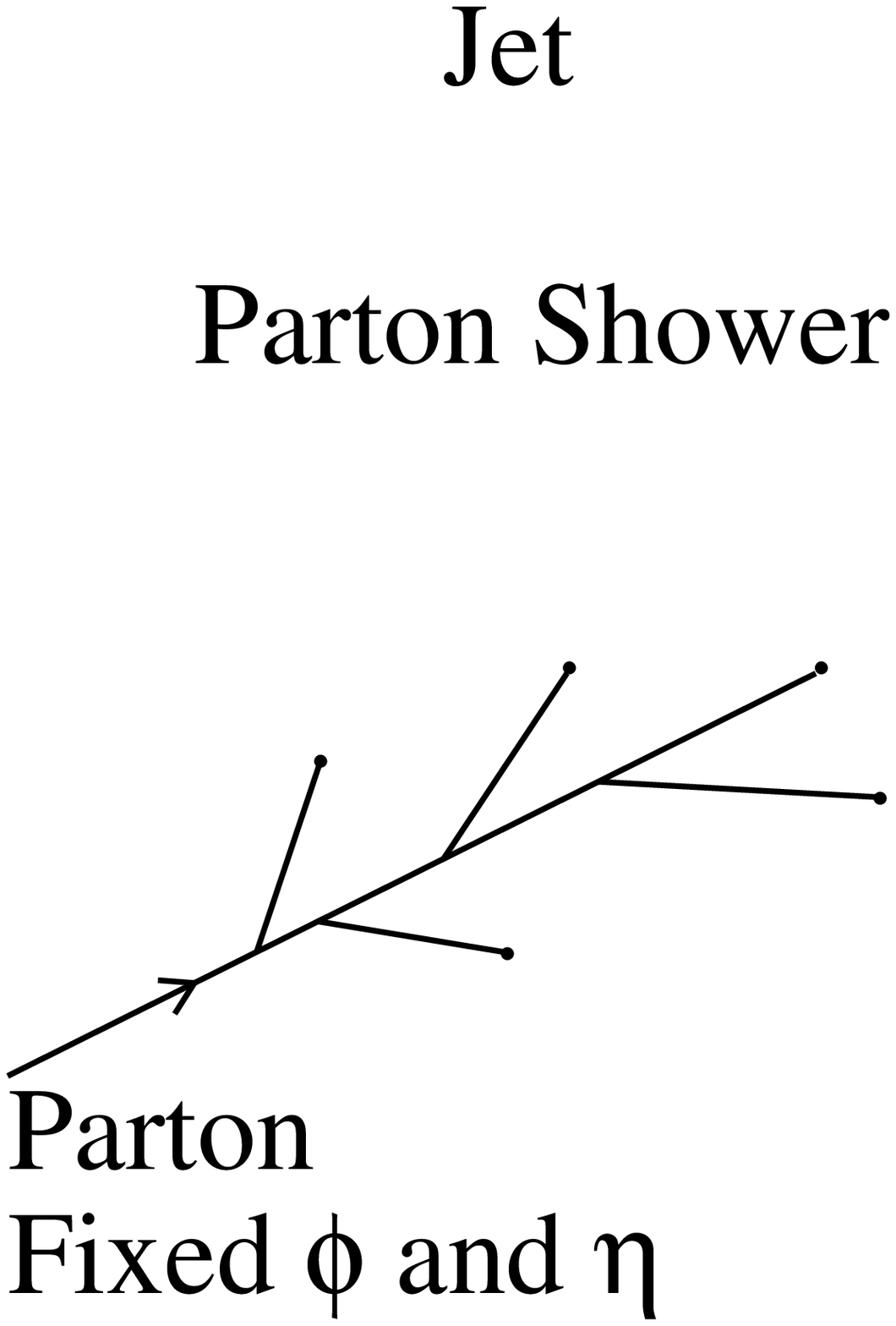}}
\end{center}
\vspace{2pt}
\caption{A jet parton shower.}
\label{fig2}
\end{figure}

\begin{figure}
\begin{center}
\mbox{
   \epsfysize 5.0in
   \epsfbox{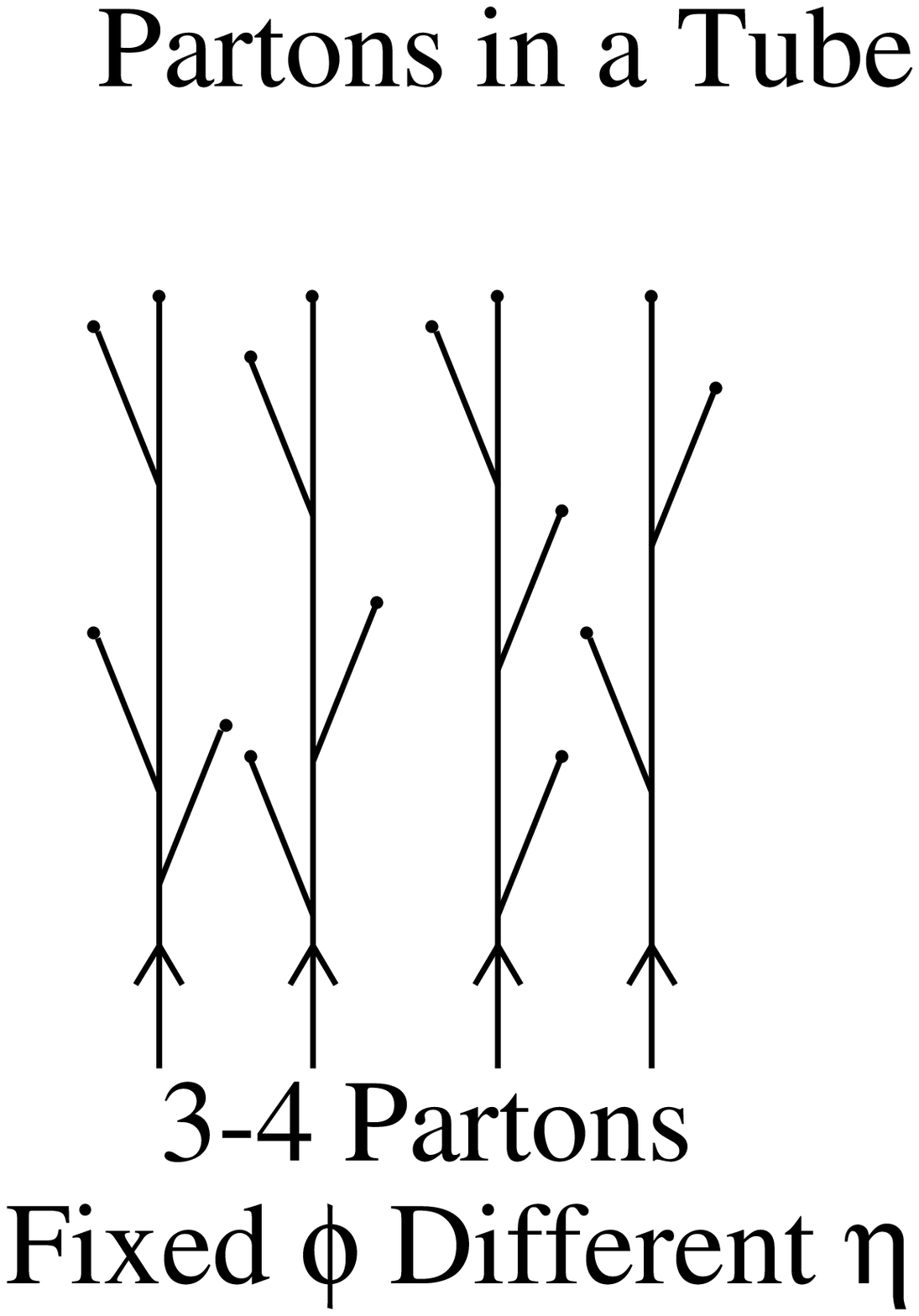}}
\end{center}
\vspace{2pt}
\caption{Each tube contains 3-4 partons as shown.}
\label{fig3}
\end{figure}

 The space momentum correlation of the blast wave provides us with a strong 
angular correlation signal. PYTHIA fragmentation functions\cite{pythia} were 
used for the tube fragmentation that generate the final state particles 
emitted from the tube. The initial transverse size of a flux tube 
$\sim$1/4 fm has expanded to the size of $\sim$2 fm at kinetic freezeout. 
Many particles that come from the surface of the fireball will have a 
$p_t$ greater than 0.8 Gev/c. The final state tube size and the Hanbury-Brown 
and Twiss (HBT) observations\cite{HBT} of pions that have a momentum range 
greater than 0.8 GeV/c are consistent both being $\sim$2 fm. A single parton 
using PYTHIA forms a jet with the parton having a fixed $\eta$ and $\phi$ 
(see Figure 2). For central events each of the twelve tubes have 3-4 partons 
per tube each at a fixed $\phi$ for a given tube. The $p_t$ distribution of 
the partons is similar to pQCD but has a suppression at high $p_t$ like the 
data. The 3-4 partons in the tube which shower using PYTHIA all have a 
different $\eta$ values but all have the same $\phi$ (see Figure 3). The 
PBM explained the high precision Au + Au central (0-10\%) collisions at 
$\sqrt{s_{NN}} =$ 200 GeV\cite{centralproduction}(the highest RHIC energy).

\section{The Correlation Function for Central Au + Au Data}

We utilize a two particle correlation function in the two dimensional (2-D)
space\footnote{$\Delta \phi = \phi_1 - \phi_2$ where $\phi$ is the azimuthal
angle of a particle measured in a clockwise direction about the beam.
$\Delta \eta = \eta_1 - \eta_2$ which is the difference of the psuedorapidity
of the pair of particles} of $\Delta \phi$ versus $\Delta \eta$. The 2-D 
total correlation function is defined as:

\begin{equation}
C(\Delta \phi,\Delta \eta)
=S(\Delta \phi,\Delta \eta)/M(\Delta \phi,\Delta \eta).
\end{equation}

Where S($\Delta \phi,\Delta \eta$) is the number of pairs at the corresponding
values of $\Delta \phi,\Delta \eta$ coming from the same event, after we have
summed over all the events. M($\Delta \phi,\Delta \eta$) is the number of pairs
at the corresponding values of $\Delta \phi,\Delta \eta$ coming from the mixed
events, after we have summed over all our created mixed events. A mixed event
pair has each of the two particles chosen from a different event. We make on
the order of ten times the number of mixed events as real events. We rescale
the number of pairs in the mixed events to be equal to the number of pairs in
the real events. This procedure implies a binning in order to deal with finite
statistics. The division by M($\Delta \phi,\Delta \eta$) for experimental data
essentially removes or drastically reduces acceptance and instrumental effects.
If the mixed pair distribution was the same as the real pair distribution 
C($\Delta \phi,\Delta \eta$) should have unit value for all of the binned 
$\Delta \phi,\Delta \eta$. In the correlations used first in this paper we 
select particles independent of its charge. The correlation of this type is 
called a Charge Independent (CI) Correlation. This difference correlation 
function has the defined property that it only depends on the differences of 
the azimuthal angle ($\Delta \phi$) and the beam angle ($\Delta \eta$) for the 
two particle pair. Thus the two dimensional difference correlation distribution
for each tube which is part of C($\Delta \phi,\Delta \eta$) is similar for 
each of the objects and will image on top of each other. We further divide the 
data (see Table I) into $p_t$ ranges (bins). 

\bf Table I. \rm The $p_t$ bins and the number of charged particles per bin 
with $| \eta |$ $<$ 1.0.

\begin{center}
\begin{tabular}{|c|r|}\hline
\multicolumn{2}{|c|}{Table I}\\ \hline
$p_t$ range & amount \\ \hline
$4.0 GeV/c - 1.1 GeV/c$ & 149 \\ \hline
$1.1 GeV/c - 0.8 GeV/c$ & 171 \\ \hline
$0.8 GeV/c - 0.65 GeV/c$ & 152 \\ \hline
$0.65 GeV/c - 0.5 GeV/c$ & 230 \\ \hline
$0.5 GeV/c - 0.4 GeV/c$ & 208 \\ \hline
$0.4 GeV/c - 0.3 GeV/c$ & 260 \\ \hline
$0.3 GeV/c - 0.2 GeV/c$ & 291 \\ \hline
\end{tabular}
\end{center}

Since we are choosing particle pairs, we choose for the first particle $p_{t1}$
which could be in one bin and for the second particle $p_{t2}$ which could be 
in another bin. Therefore binning implies a matrix of $p_{t1}$ vs $p_{t2}$. We 
have 7 bins thus there are 28 independent combination. Each combination will 
have a different number of enters. In order to take out this difference 
one uses multiplicity scaling\cite{PBME,centralitydependence}. The 
diagonal bins one scales event average of Table I. For the off diagonal 
combination one uses the product of square root of corresponding diagonal 
event averages. In Figure 4 we show the correlation function equation 1 for the
highest diagonal bin $p_t$ 4.0 to 1.1 GeV/c. Figure 5 is the smallest diagonal 
bin $p_t$ 0.3 to 0.2 GeV/c.

\begin{figure}
\begin{center}
\mbox{
   \epsfysize 6.5in
   \epsfbox{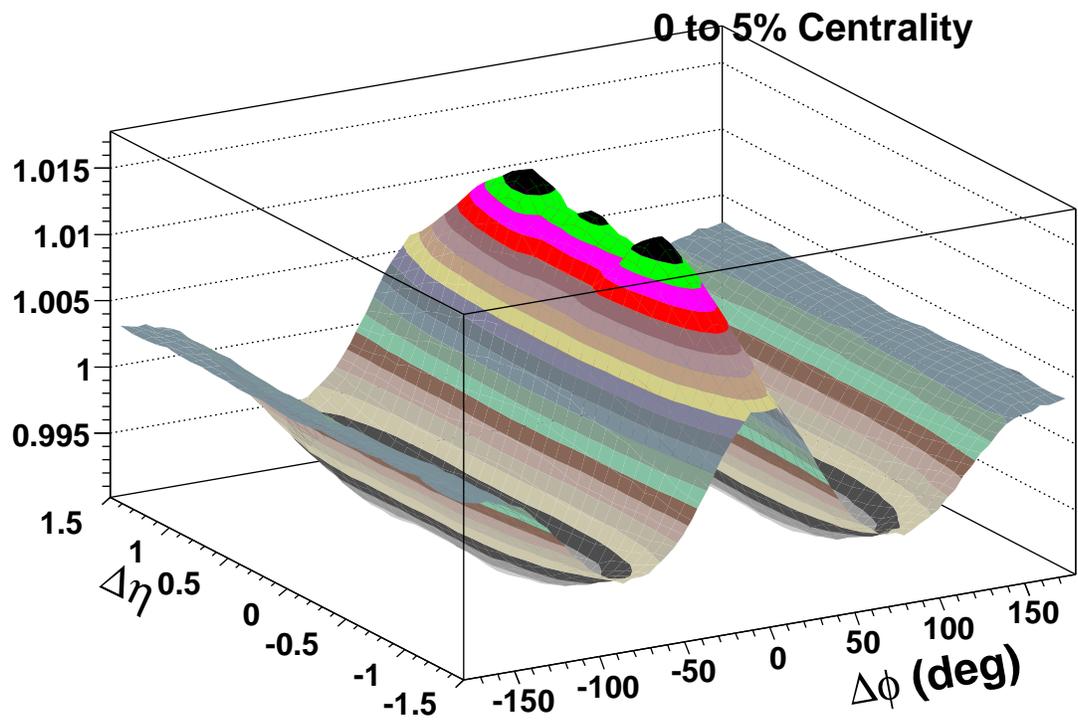}}
\end{center}
\vspace{2pt}
\caption{$\Delta \phi$ vs.$\Delta \eta$ CI correlation for the 0-5\% 
centrality bin for Au + Au collisions at $\sqrt{s_{NN}} = $ 200 GeV requiring 
both particles to be in bin 7 $p_t$ greater than 1.1 GeV/c and $p_t$ 
less than 4.0 GeV/c.}
\label{fig4}
\end{figure}

\begin{figure}
\begin{center}
\mbox{
   \epsfysize 6.5in
   \epsfbox{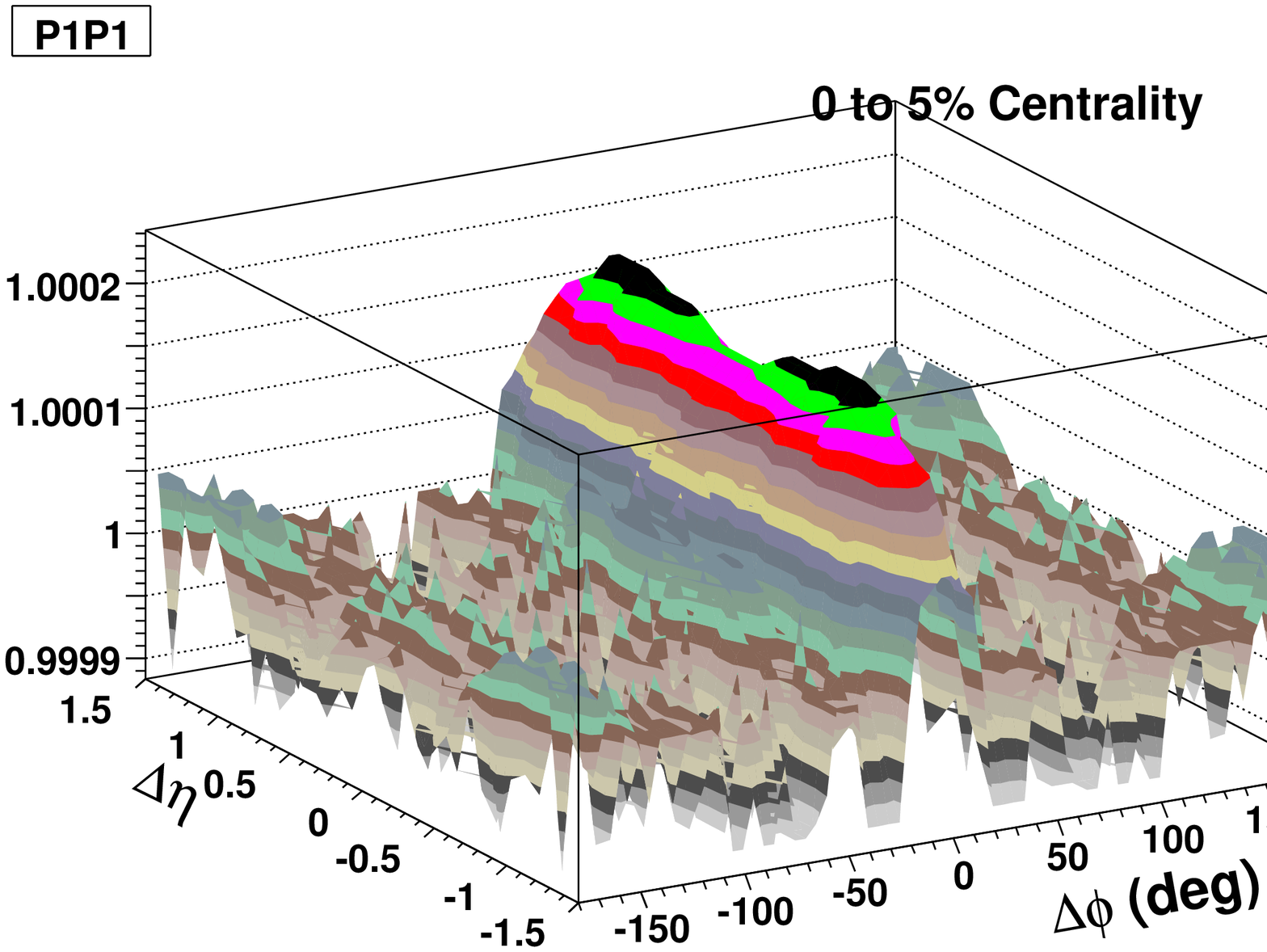}}
\end{center}
\vspace{2pt}
\caption{$\Delta \phi$ vs.$\Delta \eta$ CI correlation for the 0-5\% 
centrality bin for Au + Au collisions at $\sqrt{s_{NN}} = $ 200 GeV requiring 
both particles to be in bin 1 $p_t$ greater than 0.2 GeV/c and $p_t$ 
less than 0.3 GeV/c.}
\label{fig5}
\end{figure}

\begin{figure}
\begin{center}
\mbox{
   \epsfysize 6.5in
   \epsfbox{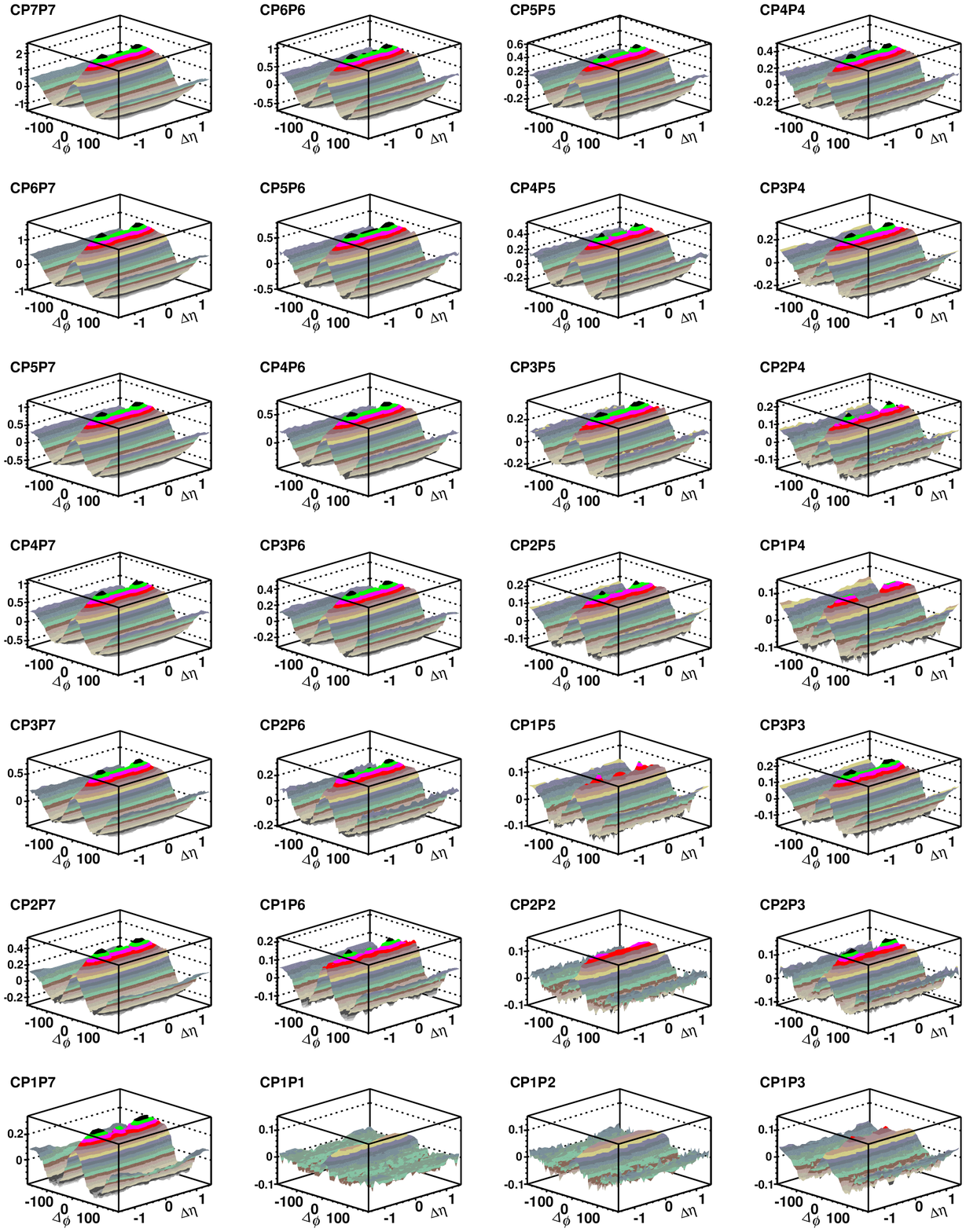}}
\end{center}
\vspace{2pt}
\caption{$\Delta \phi$ vs.$\Delta \eta$ CI correlation for the 0-5\% 
centrality bin for Au + Au collisions at $\sqrt{s_{NN}} = $ 200 GeV for 
all 28 combination of $p_{t1}$ vs $p_{t2}$ (see text).}
\label{fig6}
\end{figure}

\begin{figure}
\begin{center}
\mbox{
   \epsfysize 6.5in
   \epsfbox{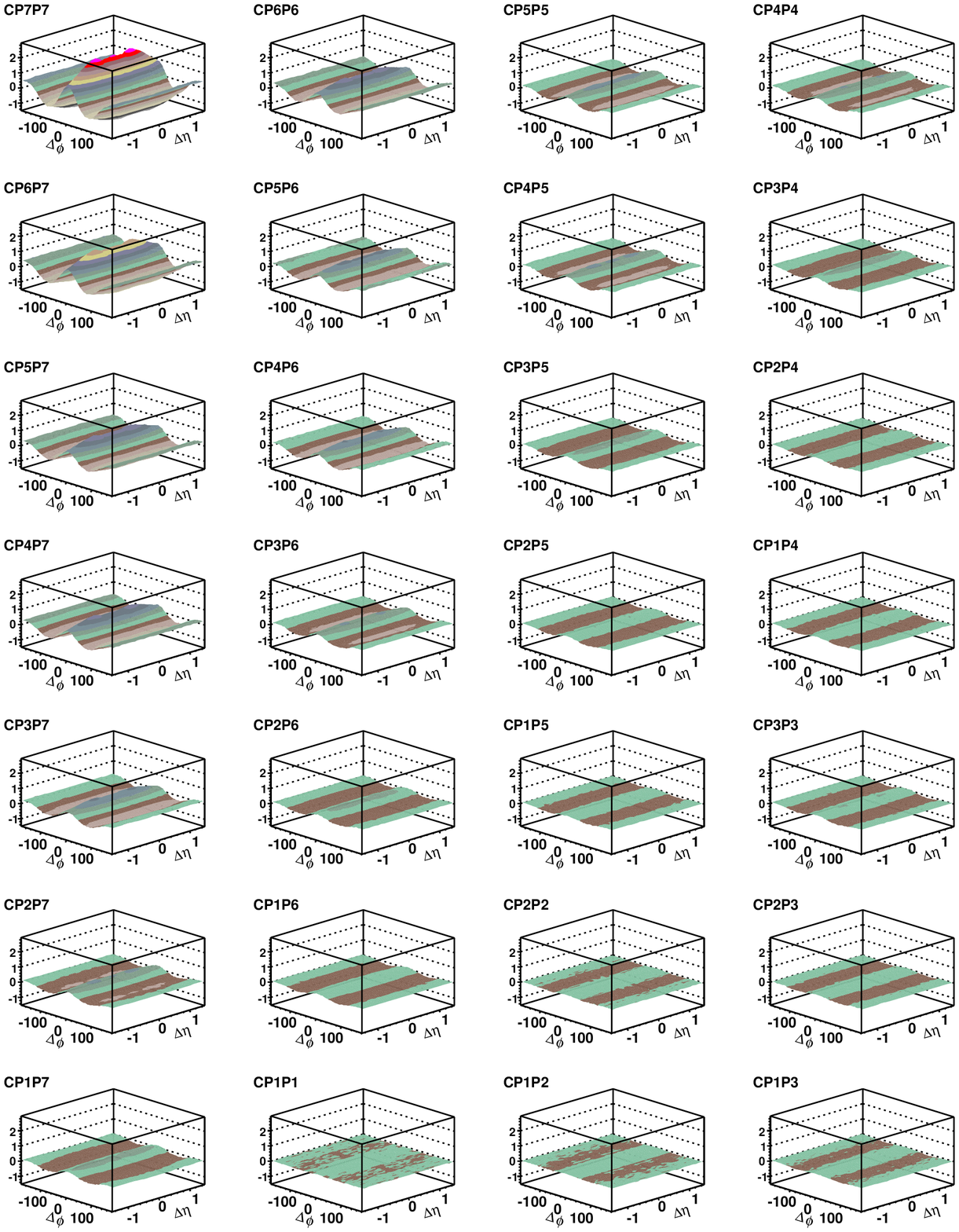}}
\end{center}
\vspace{2pt}
\caption{$\Delta \phi$ vs.$\Delta \eta$ CI correlation for the 0-5\% 
centrality bin for Au + Au collisions at $\sqrt{s_{NN}} = $ 200 GeV for 
all 28 combination of $p_{t1}$ vs $p_{t2}$ not rescaled (see text).}
\label{fig7}
\end{figure}

Once we use multiplicity scaling we can compare all 28 combination. In
Figure 9 we show all 28 plots, but the z-axis has been rescaled so that
each appear the same size. In Figure 10 we again show the 28 plots all having
the same scale. This make it easier to see how fast the correlation signals
drop off with lowering the momentum. These plots show the properties of
parton fragmentation. P1P7 has the same signal as P2P6, P3P5, and P4P4.

\clearpage

\section{The properties of the Glasma Flux Tube Model}

For the above correlations on average there was 12 final state tubes on
the surface of the fireball per central Au + Au collision. Each tube on the 
average showered into 49 charged particles. The soft uncorrelated particles 
accounted for 873 or 60\% of the charged particles. Since the tubes are
sitting on the surface of the fireball and being push outward by radial flow,
not all particles emitted from the tube will escape. Approximately one half of
the particles that are on the outward surface leave the fireball and the other 
half are absorbed by the fireball (see Figure 8). This means that 20\% of 
the charged particles come from tube emission, and 294 particles are added to 
the soft particles increasing the number to 1167 (see Table III).

\bf Table III. \rm Parameters of the GFTM for charged particles.

\begin{center}
\begin{tabular}{|c|r|r|}\hline
\multicolumn{3}{|c|}{Table III}\\ \hline
variable & amount & fluctuations \\ \hline
$tubes$ & 12 & 0  \\ \hline
$particles$& 24.5  & 5  \\ \hline
$soft$ & 1167 & 34  \\ \hline
\end{tabular}
\end{center}

The particles that emitted outward are boosted in momentum, while the inward
particles are absorbed by the fireball. Out of the initial 49 particles per
tube the lower $p_t$ particles have larger losses. In Table IV we give a
detailed account of these percentage losses and give the average number of
charged particles coming from each tube for each $p_t$ bin.

\bf Table IV. \rm Parameters of the GFTM for $p_t$ of the charged particles .

\begin{center}
\begin{tabular}{|c|r|r|}\hline
\multicolumn{3}{|c|}{Table IV}\\ \hline
$p_t$ & amount & \%survive \\ \hline
$4.0 GeV/c - 1.1 GeV/c$ & 4.2 & 100 \\ \hline
$1.1 GeV/c - 0.8 GeV/c$ & 3.8 & 76 \\ \hline
$0.8 GeV/c - 0.65 GeV/c$ & 3.2 & 65 \\ \hline
$0.65 GeV/c - 0.5 GeV/c$ & 4.2 & 54 \\ \hline
$0.5 GeV/c - 0.4 GeV/c$ & 3.2 & 43 \\ \hline
$0.4 GeV/c - 0.3 GeV/c$ & 3.3 & 35 \\ \hline
$0.3 GeV/c - 0.2 GeV/c$ & 2.6 & 25 \\ \hline
\end{tabular}
\end{center}
 
In the surface GFTM we have thermalization and hydro flow for the soft 
particles, while all the two particle angular correlations come from
the tubes on the surface. The charge particle spectrum of the GFTM is given
by a blast wave model and the direct tube fragmentation is only 20\% of this
spectrum. The initial anisotropic azimuthal distribution of flux tubes is 
transported to the final state leaving its pattern on the ring of 
final state flux tubes on the surface. This final state anisotropic flow
pattern can be decomposed in a Fourier series ($v_1$, $v_2$, $v_3$, ...).
These coefficients have been measure\cite{Alver} and have been found to be
important in central Au + Au collisions. 

\begin{figure}
\begin{center}
\mbox{
   \epsfysize 6.5in
   \epsfbox{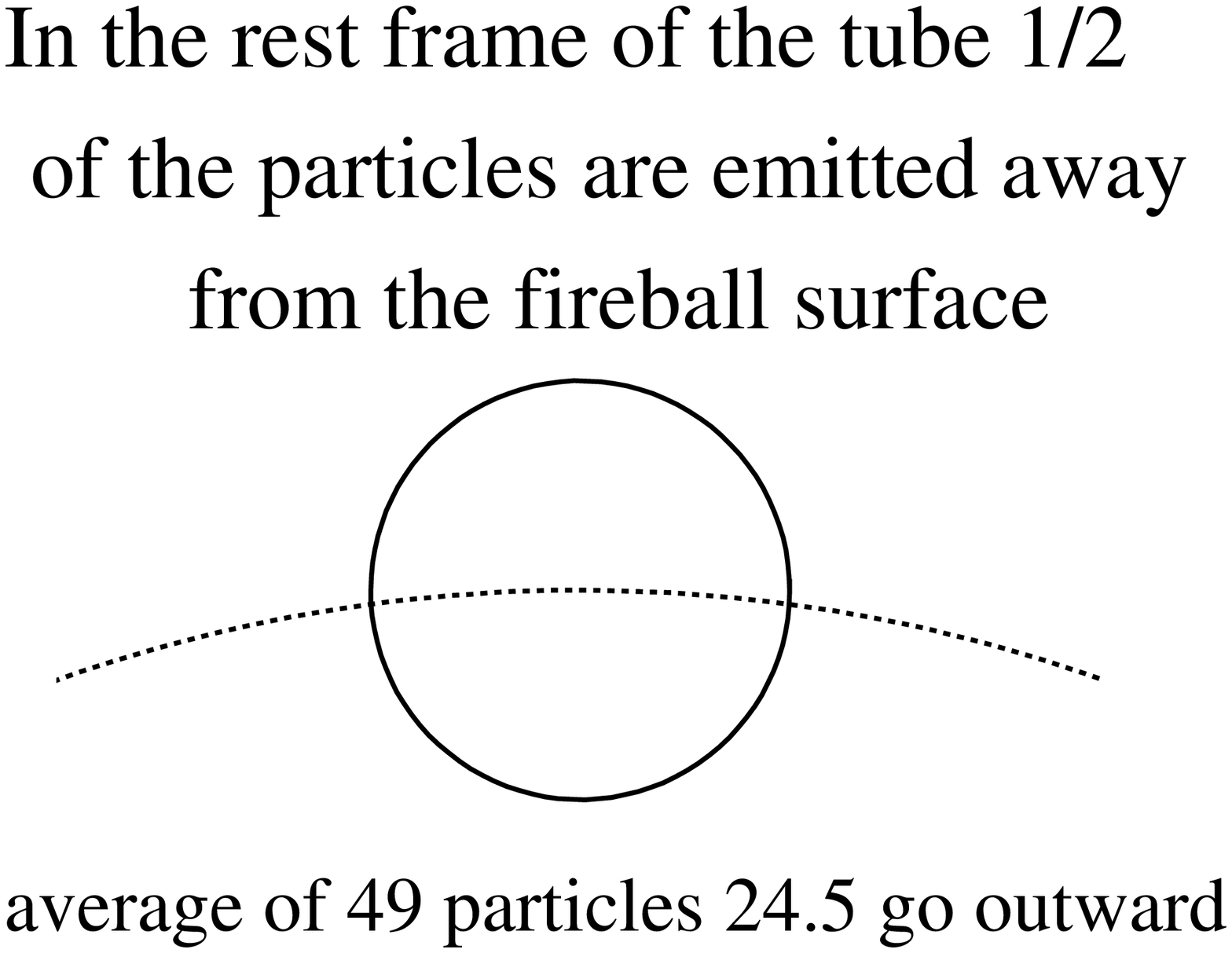}}
\end{center}
\vspace{2pt}
\caption{Each tube sits on the surface.}
\label{fig8}
\end{figure}

\section{Charge Independent Correlation compared with the Au + Au Data}

How well does the glasma flux tube model (GFTM)\cite{Dumitru}\cite{PBM}
reproduce the Au + Au central collision data at $\sqrt{s_{NN}} = $ 200 GeV.
A complete analysis for Au + Au central data 0-10\% within a $p_t$ range of 
0.8 $<$ $p_t$ $<$ 2.0 GeV/c can be found in Ref.\cite{centralproduction}.
Here we compare two of our $p_t$ ranges in a  $\Delta\eta$ range 
0.6 $<$ $\Delta\eta$ $<$ 0.9 for Au + Au central collision data 0-5\% at 
$\sqrt{s_{NN}} = $ 200 GeV. The first $p_t$ range bin is P7P7 where P7 is 1.1 
$<$ $p_t$ $<$ 4.0 Gev/c(see Figure 9). The surface flux tube correlations are 
very strong in this bin. When we let one of the particles move down in 
$p_t$ to P4 0.5 $<$ $p_t$ $<$ 0.65 Gev/c(see Figure 10) the correlation 
weakens a lot. We see that the GFTM does a good job reproducing the charge
independent data.

\begin{figure}
\begin{center}
\mbox{
   \epsfysize 6.5in
   \epsfbox{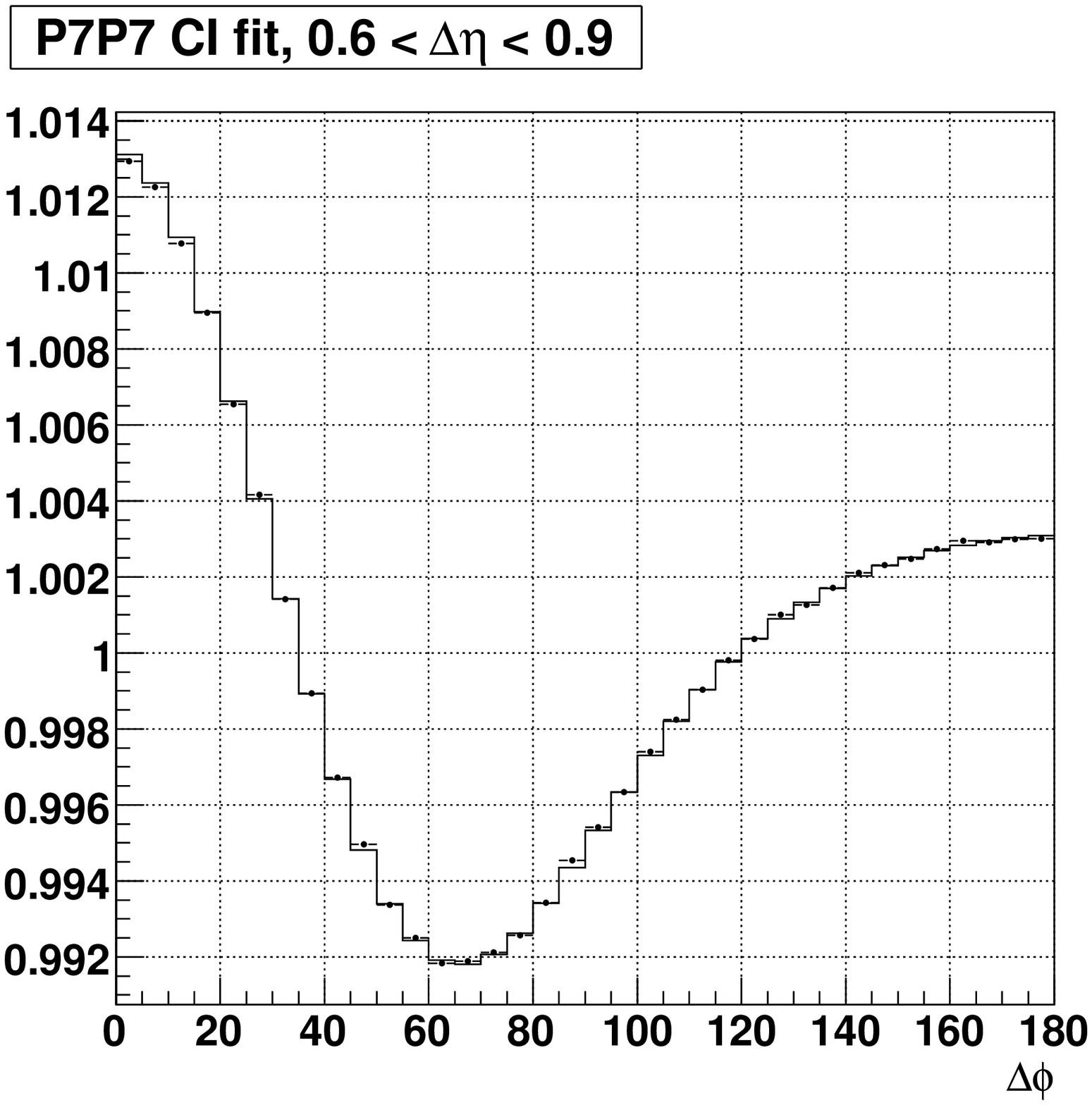}}
\end{center}
\vspace{2pt}
\caption{$\Delta \phi$ for GFTM vs Au + Au data for $\Delta\eta$ range 0.6 $<$ 
$\Delta\eta$ $<$ 0.9 Au + Au central collision data 0-5\% at $\sqrt{s_{NN}} = $ 
200 GeV requiring both particles to be in bin 7 $p_t$ greater than 1.1 GeV/c 
and $p_t$ less than 4.0 GeV/c.}
\label{fig9}
\end{figure}

\begin{figure}
\begin{center}
\mbox{
   \epsfysize 6.5in
   \epsfbox{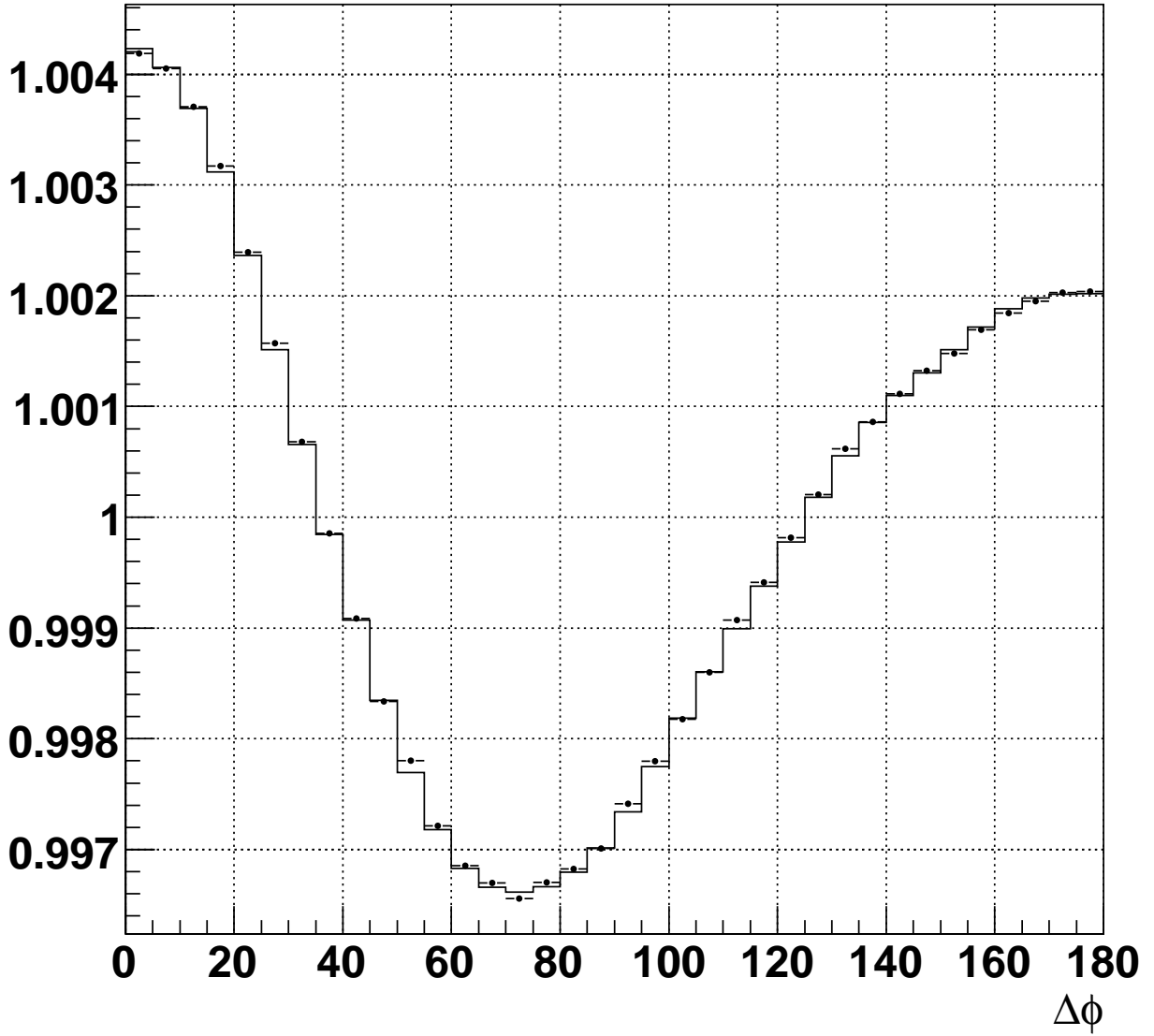}}
\end{center}
\vspace{2pt}
\caption{$\Delta \phi$ for GFTM vs Au + Au data for $\Delta\eta$ range 0.6 $<$ 
$\Delta\eta$ $<$ 0.9 Au + Au central collision data 0-5\% at $\sqrt{s_{NN}} = $ 
200 GeV requiring one particle to be in bin 7 $p_t$ greater than 1.1 GeV/c 
and $p_t$ less than 4.0 GeV/c and the second to be in bin 4 $p_t$ greater than 
0.65 eV/c and $p_t$ less than 0.8 GeV/c.}
\label{fig10}
\end{figure}

\begin{figure}
\begin{center}
\mbox{
   \epsfysize 4.2in
   \epsfbox{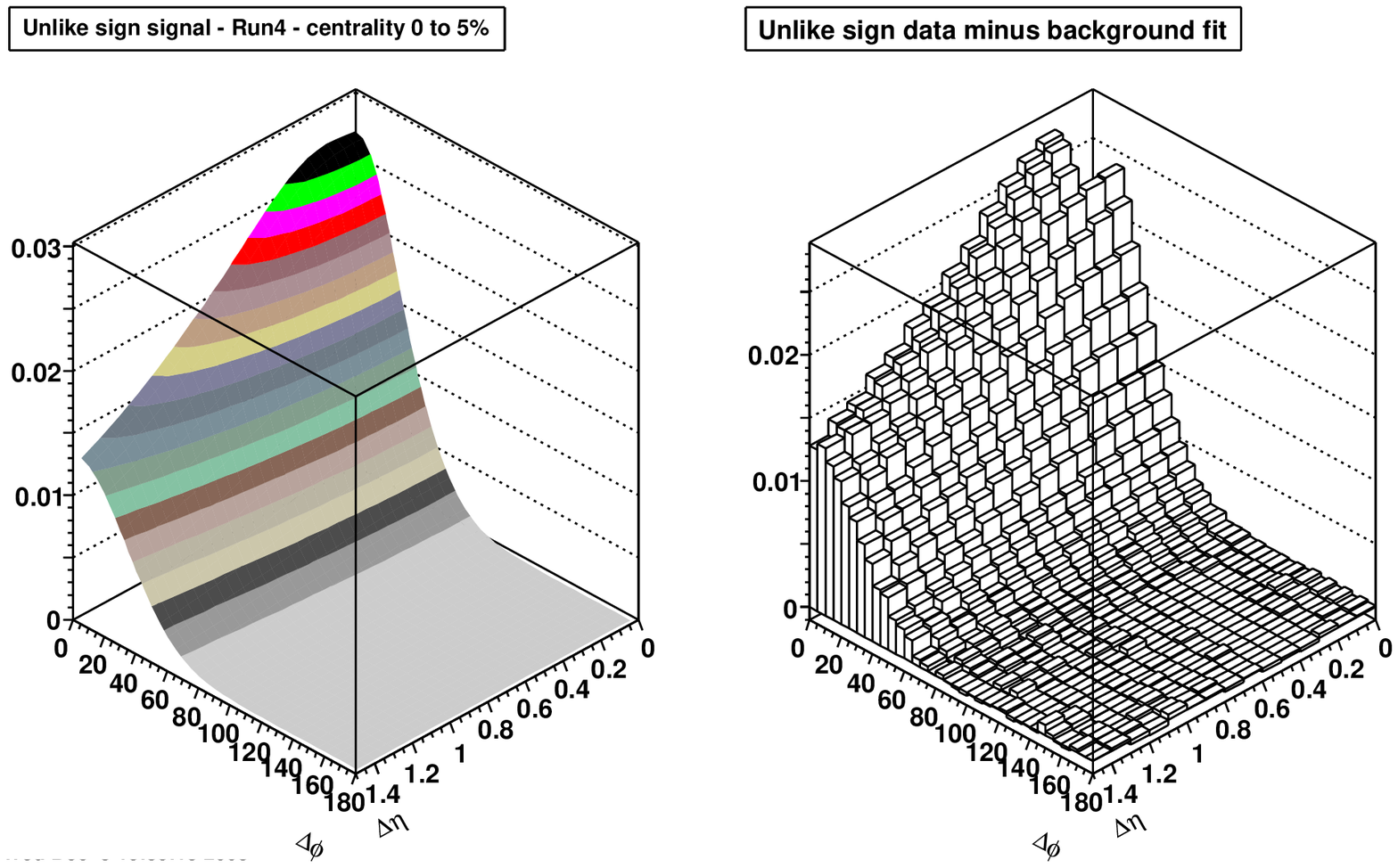}}
\end{center}
\vspace{2pt}
\caption{a)2-D perspective plot of the unlike sign pairs for a  single tube 
GFTM signal in Au + Au central collision data 0-5\% at $\sqrt{s_{NN}} = $ 200 
GeV requiring both particles to be in bin 7 $p_t$ greater than 1.1 GeV/c 
and $p_t$ less than 4.0 GeV/c. b)2-D perspective plot of the unlike sign pairs
single tube data signal(see text) in Au + Au central collision data 0-5\% at 
$\sqrt{s_{NN}} = $ 200 GeV requiring both particles to be in bin 7 $p_t$ 
greater than 1.1 GeV/c and $p_t$ less than 4.0 GeV/c.}
\label{fig11}
\end{figure}

\begin{figure}
\begin{center}
\mbox{
   \epsfysize 4.2in
   \epsfbox{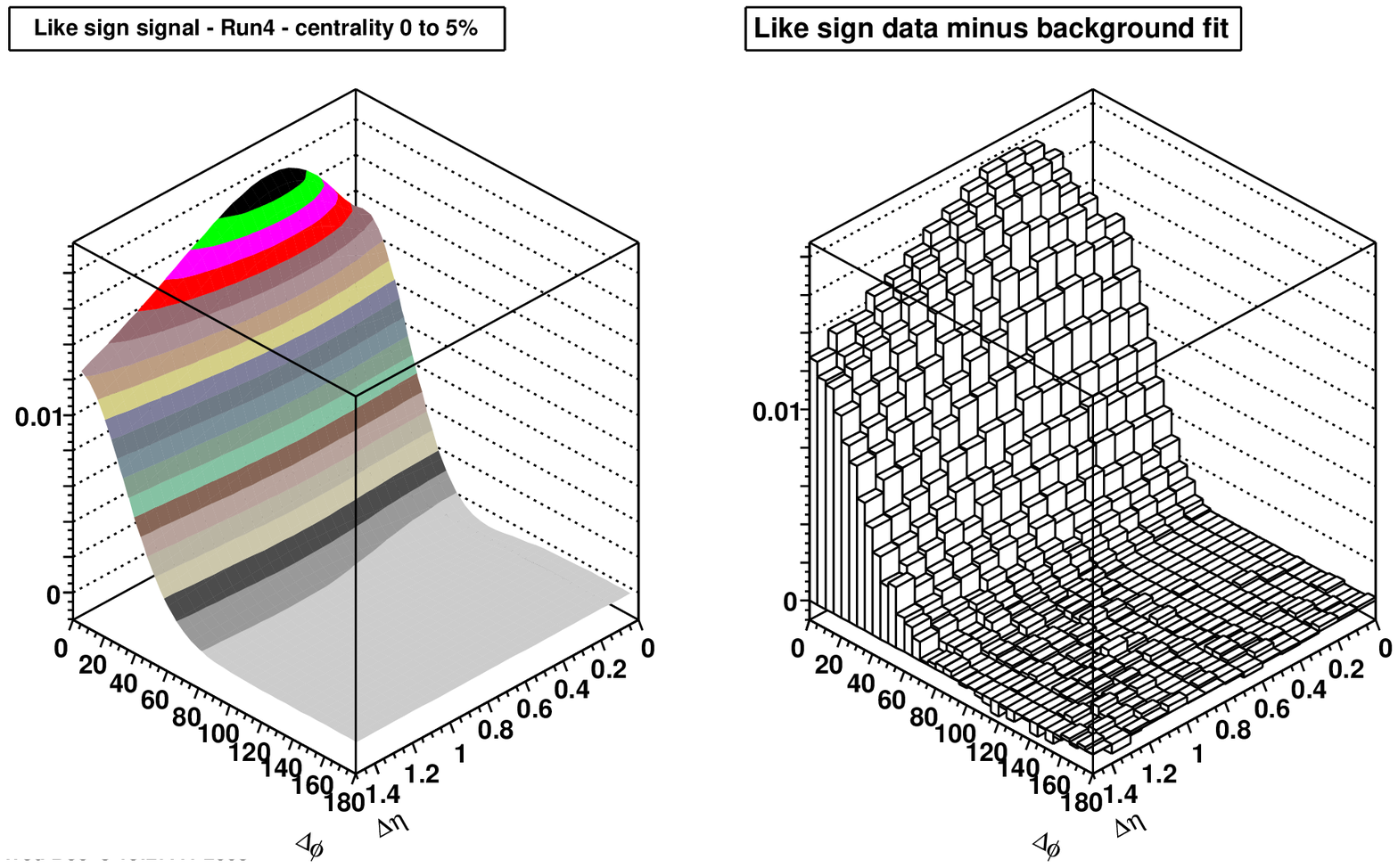}}
\end{center}
\vspace{2pt}
\caption{a)2-D perspective plot of the like sign pairs for a  single tube 
GFTM signal in Au + Au central collision data 0-5\% at $\sqrt{s_{NN}} = $ 200 
GeV requiring both particles to be in bin 7 $p_t$ greater than 1.1 GeV/c 
and $p_t$ less than 4.0 GeV/c. b)2-D perspective plot of the like sign pairs
single tube data signal(see text) in Au + Au central collision data 0-5\% at 
$\sqrt{s_{NN}} = $ 200 GeV requiring both particles to be in bin 7 $p_t$ 
greater than 1.1 GeV/c and $p_t$ less than 4.0 GeV/c.}
\label{fig12}
\end{figure}

\begin{figure}
\begin{center}
\mbox{
   \epsfysize 7.4in
   \epsfbox{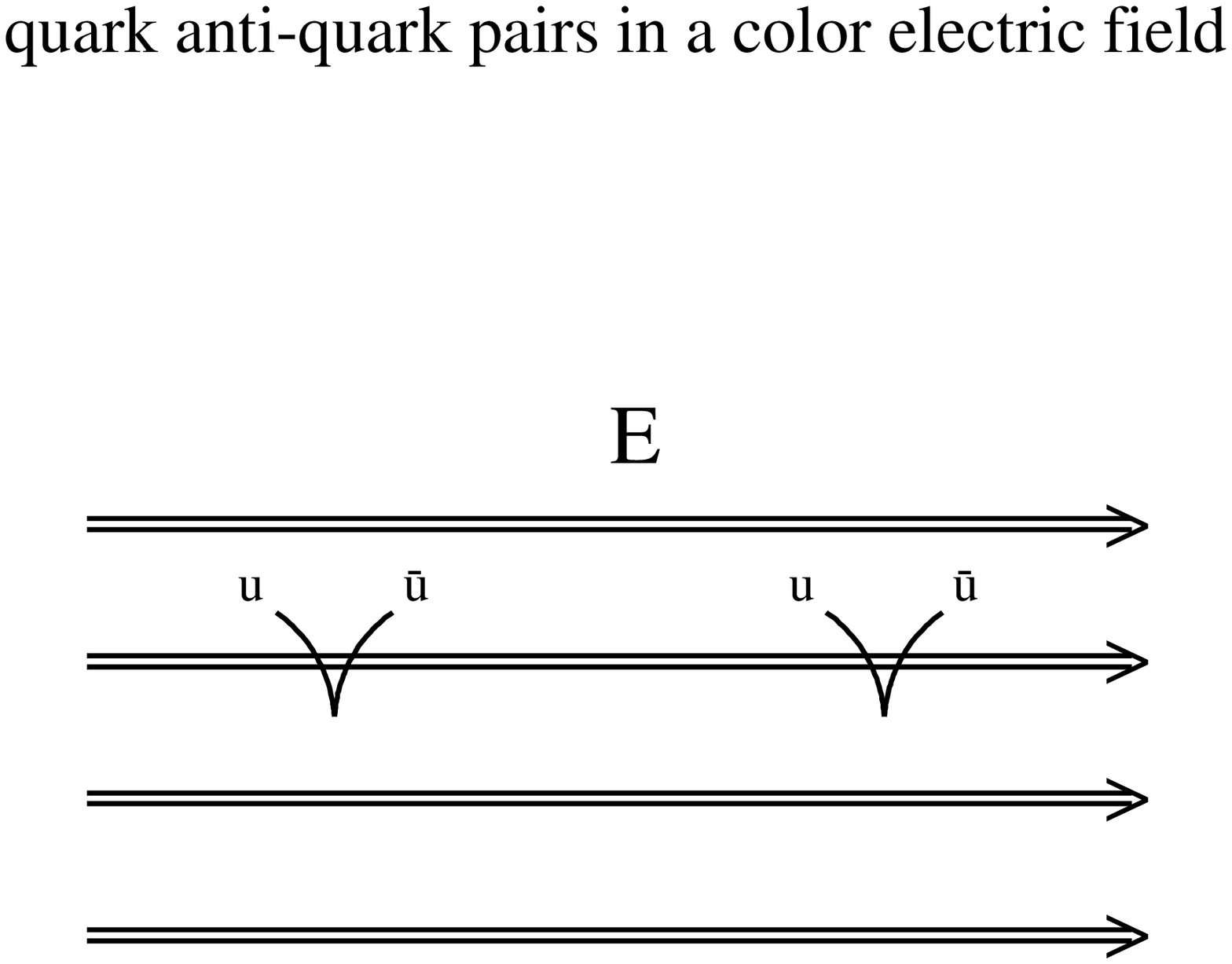}}
\end{center}
\caption{Inside each flux tube a large color electric field causes pairs of
color and anti-color up quarks to be aligned in this field. The up quark ends
up in a $\pi^+$ and the anti-up quark ends up in a $\pi^-$.}
\label{fig13}
\end{figure}

\section{Charge dependent correlations of the signal coming from a single flux 
tube}
  
In this section we are going to explore the detailed shape of the correlations
coming from a single flux tube. We will follow the method used in 
Refs.\cite{PBM,centralproduction,PBME,centralitydependence}. We can display
the correlations coming from a single flux tube if we form S of equation 1,
where S($\Delta \phi,\Delta \eta$) is the number of pairs at the corresponding
values of $\Delta \phi,\Delta \eta$ coming from the same flux tube in each 
event. We then sum S over all the events. M($\Delta \phi,\Delta \eta$) 
remains the number of pairs at the corresponding values of 
$\Delta \phi,\Delta \eta$ coming from the mixed events, after we have summed 
over all our created mixed events. Another way to form is same correlation is
to form S of equation 1 coming from all pairs except those from the
same flux tube in each event(we call this the background). We then subtract 
this S(background) from S made out of all the pairs. If we want to display the 
single flux tube correlation of the data, we subtract S(background) from the 
S generated by the data. 

In order to explore the charge dependence, we separate pairs into like sign 
pairs and unlike sign pairs. We will show how well GFTM describes surface flux 
tubes by looking at P7P7 charge dependent correlation. In Figure 11 we show the
unlike sign pair correlation of the single flux tube and the background 
subtracted data. In Figure 12 we show the same for like sign pairs.
It should be noted that when we compare the data there are some pair wise
track cuts. For Figure 11 unlike sign pairs with a $\Delta\eta$ $<$ 0.1
there is a $\Delta\phi$ $<$ $20^0$ cut and 0.1 $<$ $\Delta\eta$ $<$ 0.2 
there is a $\Delta\phi$ $<$ $10^0$ cut. These cuts remove track merging and 
the Coulomb effect. For Figure 12 like sign pairs with a $\Delta\eta$ $<$ 0.2
there is a $\Delta\phi$ $<$ $5^0$ cut. These cuts remove track merging and the 
HBT effect. 

We see in both Figure 11 and Figure 12, that the correlation from each pair
choice is well described by GFTM. We see at small $\Delta \phi,\Delta \eta$
there is a local bump for unlike sign pairs, while there is negative bump or
dip for like sign pairs. This dip or hole torn is caused by QCD field shown
in Figure 13. This field focus unlike charge pairs into the same 
$\Delta \phi,\Delta \eta$ and defocus like charge pairs out of this same
region.

\section{Strong CP violations or Chern-Simons topological charge}

\subsection{The Source $F \widetilde F$.}

The strong CP problem remains one of the most outstanding puzzles of the
Standard Model. Even though several possible solutions have been put forward
it is not clear why CP invariance is respected by the strong interaction.
It was shown however through a theorem by Vafa-Witten\cite{Witten} that the
true ground state of QCD cannot break CP. 

The part of the QCD Lagrangian that breaks CP is related to the gluon-gluon 
interaction term $F \widetilde F$. The part of $F \widetilde F$ related 
to CP violations can be separated into a separate term which then 
can be varied by multiplying this term by a parameter called $\theta$. 
For the true QCD ground state $\theta$ = 0 (Vafa-Witten theorem). 
In the vicinity of the deconfined QCD vacuum metastable 
domains\cite{Tytgat} with $\theta$ non-zero could exist and not contradict the
Vafa-Witten theorem since it is not the QCD ground state. The metastable 
domains CP phenomenon would manifest itself in specific correlations of 
pion momenta\cite{Tytgat,Pisarski}.

\subsection{Pionic measures of CP violation.}

The glasma flux tube model (GFTM)\cite{Dumitru} considers the 
wave functions of the incoming projectiles, form sheets of CGC\cite{CGC}
at high energies that collide, interact, and evolve into high intensity color 
electric and magnetic fields. This collection of primordial fields is 
the Glasma\cite{Lappi,Gelis}. Initially the Glasma is composed of 
only rapidity independent longitudinal (along the beam axis) 
color electric and magnetic fields. These longitudinal 
color electric and magnetic fields generate topological 
Chern-Simons charge\cite{Simons} through the $F \widetilde F$ 
term and becomes a source of CP violation. How much of these 
longitudinal color electric and magnetic fields are still present in the 
surface flux tubes when they have been pushed by the blast wave is a 
speculation of this paper for measuring strong CP violation? The color 
electric field which points along the flux tube axis causes an up quark to be
accelerated in one direction along the beam axis, while the anti-up quark 
is accelerated in the other direction. So when a pair of quarks and 
anti-quarks are formed they separate along the beam axis leading to 
a separated $\pi^+$ $\pi^-$ pair along this axis(see Figure 13). The color 
magnetic field which also points along the flux tube axis (which is parallel 
to the beam axis) causes an up quark to rotate around the flux tube axis 
in one direction, while the anti-up quark is rotated in the other direction. 
So when a pair of quarks and anti-quarks are formed they will pick up or lose 
transverse momentum(see Figure 14). These changes in $p_t$ will be transmitted 
to the $\pi^+$ $\pi^-$ pairs.

\begin{figure}
\begin{center}
\mbox{
   \epsfysize 7.4in
   \epsfbox{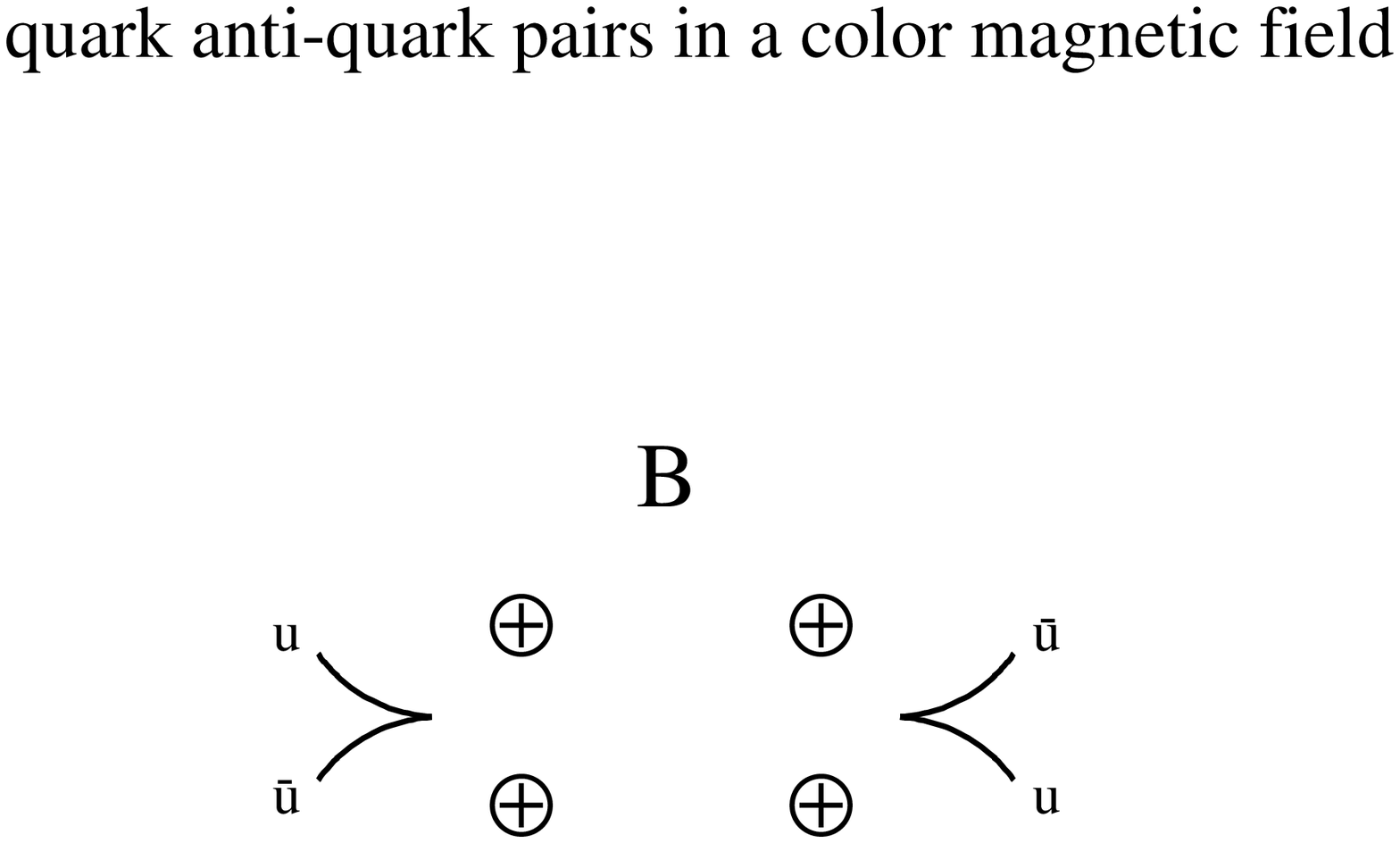}}
\end{center}
\caption{Inside each flux tube a large color magnetic field causes pairs of
color and anti-color up quarks to be pushed away from this field causing
a difference in $P_t$ between quark and anti quark. The up quark 
ends up in a $\pi^+$ and the anti-up quark ends up in a $\pi^-$.}
\label{fig14}
\end{figure}

To represent the color electric field effect we assume as the first step for
the results being shown in this paper; that we generate tubes which have an 
added boost of 100 MeV/c to the quarks in the longitudinal momentum which 
represents the color electric effect. The $\pi^+$ and the $\pi^-$ which form a 
pair are boosted in opposite directions along the beam axis. In a given tube 
all the boosts are the same but vary in direction from tube to tube. For
the color magnetic field effect, we give 100 MeV/c boosts to the transverse 
momentum in an opposite way to the $\pi^+$ and the $\pi^-$ which form a pair. 
For each pair on one side of a tube one of the charged pions boost is 
increased and the other is decreased. While on the other side of the tube 
the pion for which the boost was increased is now decreased and the pion 
which was decreased is now increased by the 100 MeV/c. All pairs for a 
given tube are treated in the same way however each tube is random 
on the sign of the pion which is chosen to be boosted on a given side. 
This addition to our model is used in the simulations of the following 
sections.

\section{Color electric field pionic measure and data.}

Above we saw that pairs of positive and negative pions should show a charge 
separation along the beam axis due to a boost in longitudinal momentum
caused by the color electric field. A measure of this separation should be a 
difference in the pseudorapidity ($\Delta \eta$) of the opposite sign pairs. 
This $\Delta \eta$ measure has a well define sign if we defined this 
difference measuring from the $\pi^-$ to the $\pi^+$. In order to form a 
correlation we must pick two pairs for comparison. The pairs have to come from 
the same tube (a final state of an expanded flux tube) since pairs 
originating from different tubes will not show this correlation. 
Therefore we require the $\pi^+$ and the $\pi^-$ differ by $20^\circ$ 
or less in $\phi$. Let us call the first pair $\Delta \eta_1$. The 
next pair $\Delta \eta_2$ (having the same $\phi$ requirement) has to also lie 
inside the same tube to show this correlation. Thus we require that there is 
only $10^\circ$ between the average $\phi$ of each pair. This implies that 
at most in $\phi$ no two pions can differ by more than $30^\circ$. In 
Figure 15 we show two pairs which would fall into the above cuts. 
$\Delta \eta_1$ and $\Delta \eta_2$ are positive in Figure 15 but if we 
would interchange the $\pi^+$ and $\pi^-$ on either pair the value of its 
$\Delta \eta$ would change sign. Finally the mean value shown on Figure 15 
is the mid-point between the $\pi^+$ and the $\pi^-$ where one really uses 
the vector sum of the $\pi^+$ and the $\pi^-$ which moves this point toward 
the harder pion.

Considering the above cuts we defined a correlation function where we combine 
pairs each having a $\Delta \eta$. Our variable is related to the sum of the 
absolute values of the individual $\Delta \eta$'s 
($\vert \Delta \eta_1 \vert +\vert \Delta \eta_2 \vert$). We assign a sign
to this sum such that if the sign of the individual $\Delta \eta$'s are
the same it is a plus sign, while if they are different it is a minus sign.
For the flux tube the color electric field extends over a large pseudorapidity
range therefore let us consider the separation of pairs 
$\vert \Delta \eta \vert$ greater than 0.9. For the numerator of the
correlation function we consider all combination of unique pairs
(sign ($\vert \Delta \eta_1 \vert +\vert \Delta \eta_2 \vert$)) from a given 
central Au + Au event divided by a mixed event denominator created from pairs
in different events. We determine the rescale of the mixed event denominator by
considering the number of pairs of pairs for the case $\vert \Delta \phi \vert$
lying between $50^\circ$ and $60^\circ$ for events and mixed events so that the
overall ratio of this sample numerator to denominator is 1. By picking 
$50^\circ$ $<$ $\vert \Delta \phi \vert$ $<$ $60^\circ$ we make sure we are
not choosing pairs from the same tube. For a simpler notation let 
(sign ($\vert \Delta \eta_1 \vert +\vert \Delta \eta_2 \vert$)) =
$\Delta \eta_1 + \Delta \eta_2$ which varies from -4 to +4 since we have an
overall $\eta$ acceptance -1 to +1 (for the STAR TPC detector for which we 
calculated). The value being near $\pm$ 4 can happen when one has a hard 
pion with $p_t$ of 4 GeV/c (upper cut) at $\eta$ = 1 with a soft pion $p_t$ 
of 0.8 GeV/c (lower cut) at $\eta$ = -1 combined with another pair; 
a hard pion with $p_t$ of 4 GeV/c at $\eta$ = -1 with a soft pion with
$p_t$ of 0.8 GeV/c at $\eta$ = 1.

In Figure 16 we show the correlation function of opposite sign 
charge-particle-pairs paired and binned by the variable 
$\Delta \eta_1 + \Delta \eta_2$ with a cut $\vert \Delta \eta \vert$ 
greater than 0.9 between the vector sums
of the two pairs.  The events are generated by the PBM\cite{PBM} and are 
charged particles of 0.8 $<$ $p_t$ $<$ 4.0 GeV/c, and $\vert \eta \vert$ $<$ 
1, from Au + Au central (0-10\%) collisions at $\sqrt{s_{NN}} =$ 200 GeV. 
Since we select pairs of pairs which are near each other in $\phi$ they all 
together pick up the tube correlation and thus these pairs of pairs overall 
show about a 4\% correlation. In the 1.0 
$<$ $\Delta \eta_1 + \Delta \eta_2$ $<$ 2.0 region the correlation is 0.5\% 
larger than the -2.0 $<$ $\Delta \eta_1 + \Delta \eta_2$ $<$ -1.0 region. This 
means there are more pairs of pairs aligned in the same direction compared to 
pairs of pairs not aligned. This alignment is what is predicted by the color 
electric field effect presented above. In fact if one has plus minus pairs all
aligned in the same direction and spread across a pseudorapidity range,
locally at any place in the pseudorapidity range one would observe
an increase of unlike sign charge pairs compared to like sign charge pairs.
At small $\Delta \eta$ unlike sign charge pairs are much larger than like
sign charge pairs in both the PBM and the data which agree. See Figures. 10, 11,
and 14 of Ref.\cite{PBM}. Figure 11 compares the total correlation for 
unlike-sign charge pairs and like-sign charge pairs in the precision STAR
central production experiment for Au + Au central (0-10\%) collisions at 
$\sqrt{s_{NN}}$=200 GeV, in the transverse momentum range 0.8 $<$ $p_t$ $<$ 
2.0 GeV/c\cite{centralproduction}. The unlike-sign charge pairs are clearly 
larger in the region near $\Delta \phi$ = $\Delta \eta$ = 0.0. The increased
correlation of the unlike-sign pairs is 0.8\% $\pm$ 0.002\%. Figure 10 
shows that the PBM fit to these data gives the same results. Figure 14 shows
that the CD = unlike-sign charge pairs minus like-sign charge pairs is positive
for the experimental analysis. Therefore the unlike-sign charge pairs are
considerably larger than the like-sign charge pairs. In fact
this effect is so large and the alignment is so great that when one adds
the unlike and like sign charge pairs correlations together there is still
a dip at small $\Delta \eta$ and $\Delta \phi$ see Figure 4 and Figure 5 of this
paper. The statistical significance of this dip in the two high precision 
experiments done independently from different data sets gathered 2 years 
apart\cite{centralproduction,centralitydependence} is huge. It would require
a fluctuation of $14\sigma$ to remove the dip. This dip is also not due to any 
systematic error since both of the just cited precision experiments carefully 
investigated that possibility; and found no evidence to challenge the reality 
of this dip. This highly significant dip ($\sim$$14\sigma$) means that 
like-sign pairs are removed as one approaches the region $\Delta \eta$ = 
$\Delta \phi$ = 0.0. Thus this is very strong evidence for the predicted 
effect of the color electric field. 

Let us now apply the above analysis to 0-10\% central Au + Au collisions at 
$\sqrt{s_{NN}} =$ 200 GeV. In figure 17 we show that aligned pairs positive
$\Delta \eta_1 + \Delta \eta_2$ is 0.2\% greater than not aligned pairs
negative $\Delta \eta_1 + \Delta \eta_2$. This effect is about half the value
we had simulated, but shows that this alignment effect is present in the data.
We can obtain a measure of how significant this difference is by forming a
ratio of not aligned(unlike) over aligned(like)(see Figure 18). If there was 
no effect the ratio would be a straight line at 1.0. We see that there is a 
dip of 0.2\% at a $\Delta \eta_1 + \Delta \eta_2$ = 1.4 with an average value
shift of 0.14\% over the expected range.

\begin{figure}
\begin{center}
\mbox{
   \epsfysize 5.4in
   \epsfbox{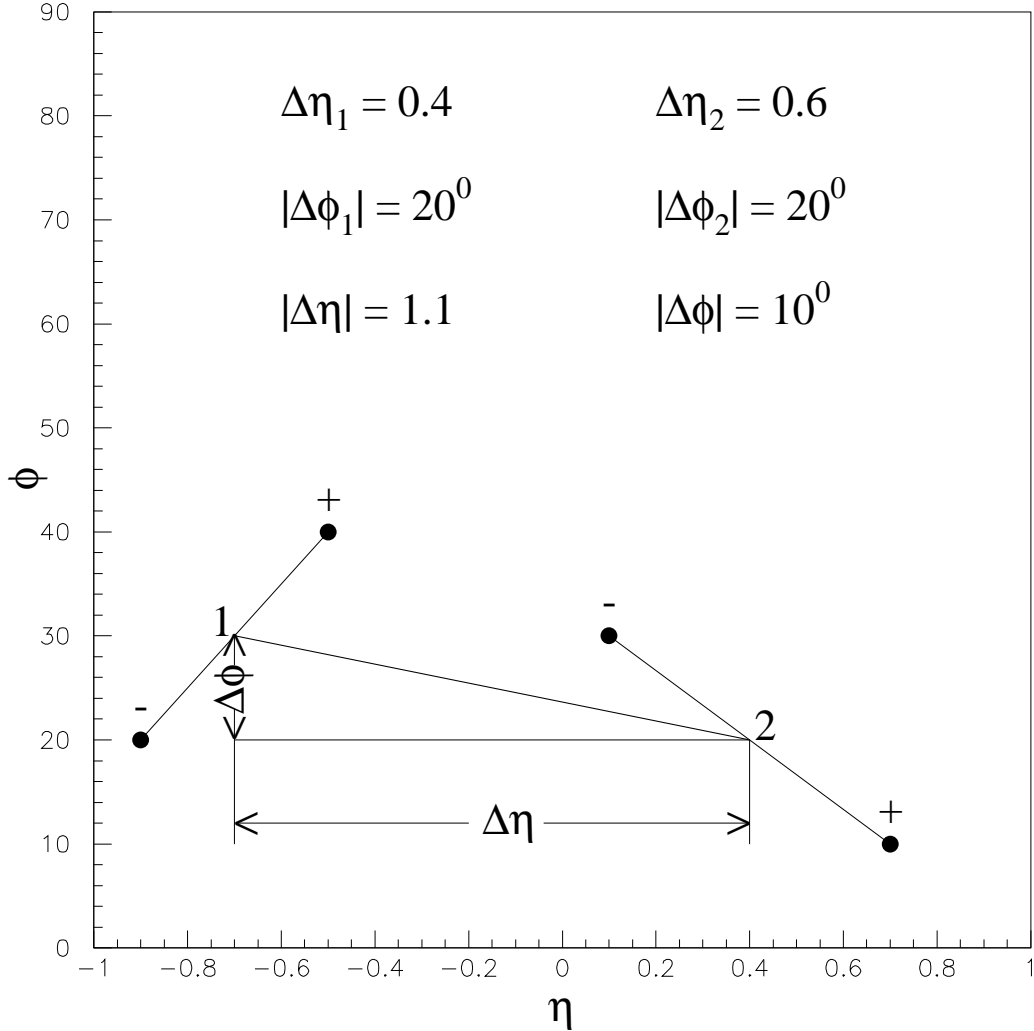}}
\end{center}
\caption{Pairs of pairs selected for forming a correlation due 
to the boost in longitudinal momentum caused by the color electric field. The 
largest separation we allow in $\phi$ for plus minus pair 1 is $\Delta \phi_1$ 
= $20^0$. This assures the pair is in the same tube. The same is true for 
pair 2 so that it will also be contained in this same tube. The mid-point for
pair 1 and 2 represents the vector sum of pair 1 and 2
which moves toward the harder particle when the momenta differ. These
mid-points cannot be separated by more than $10^0$ in $\Delta \phi$ in 
order to keep all four particles inside the same tube since the correlation
function is entirely generated within the same tube. The 
$\Delta \eta$ measure is the angle between the vector sum 1 compared to the 
vector sum 2 along the beam axis (for this case $\vert \Delta \eta \vert$ = 
1.1). The positive sign for $\Delta \eta_1$ comes from the fact that one moves 
in a positive $\eta$ direction from negative to positive. The same is true for 
$\Delta \eta_2$. If we would interchange the charge of the particles of the 
pairs the sign would change.}
\label{fig15}
\end{figure}

\begin{figure}
\begin{center}
\mbox{
   \epsfysize 5.4in
   \epsfbox{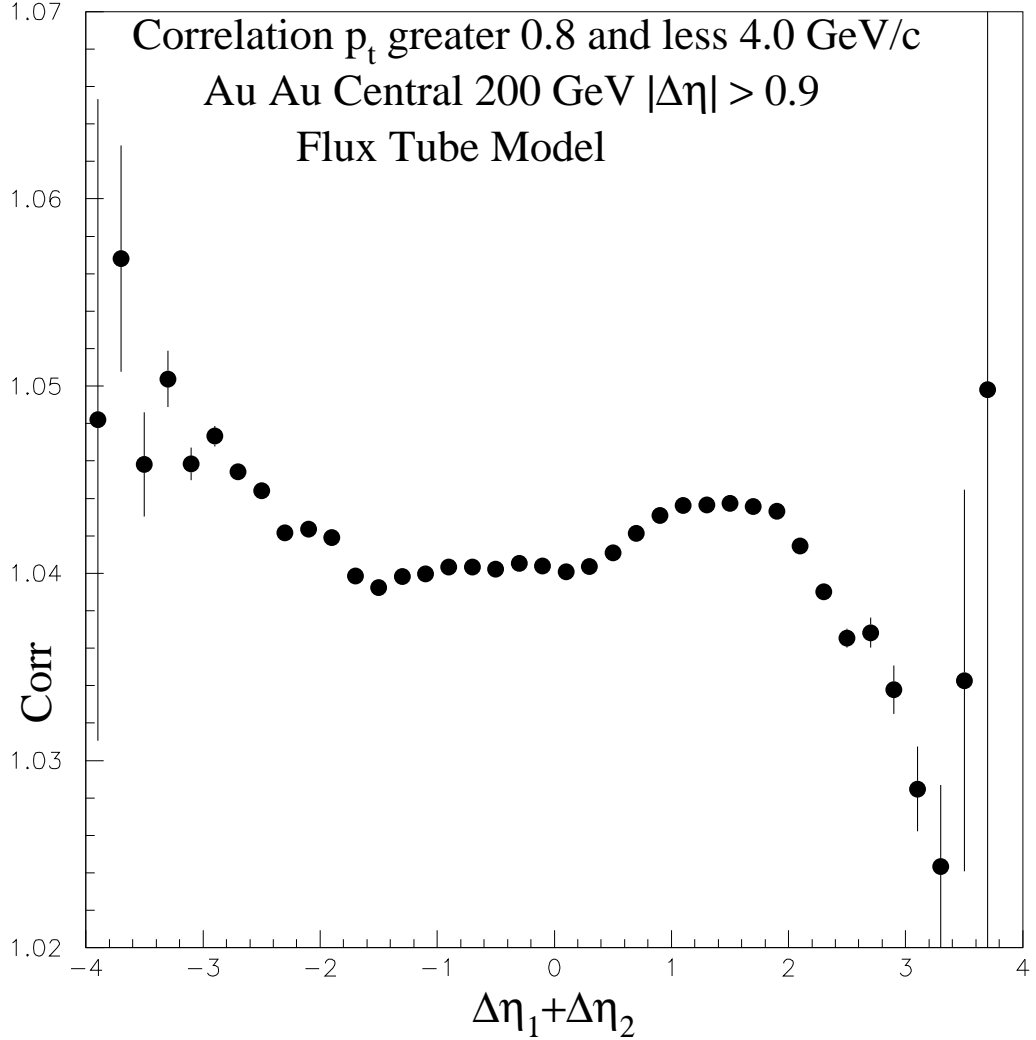}}
\end{center}
\caption{ Correlation function of pairs formed to exhibit the 
effects of longitudinal momentum boosts by the color electric field as defined 
in the text. This correlation shows that there are more aligned pairs 
(correlation is larger by $\sim$0.5\% between 1.0 $<$ $\Delta \eta_1 + 
\Delta \eta_2$ $<$ 2.0 compared to between -2.0 $<$ 
$\Delta \eta_1 + \Delta \eta_2$ $<$ -1.0). $\Delta \eta_1 + \Delta \eta_2$ 
is equal to $\vert \Delta \eta_1 \vert +\vert \Delta \eta_2 \vert$.
As explained in the text this means there are more pairs of pairs aligned 
in the same direction compared to pairs not aligned as predicted by the 
color electric field. The signs and more detail are also explained in the 
text.}
\label{fig16}
\end{figure}

\begin{figure}
\begin{center}
\mbox{
   \epsfysize 7.4in
   \epsfbox{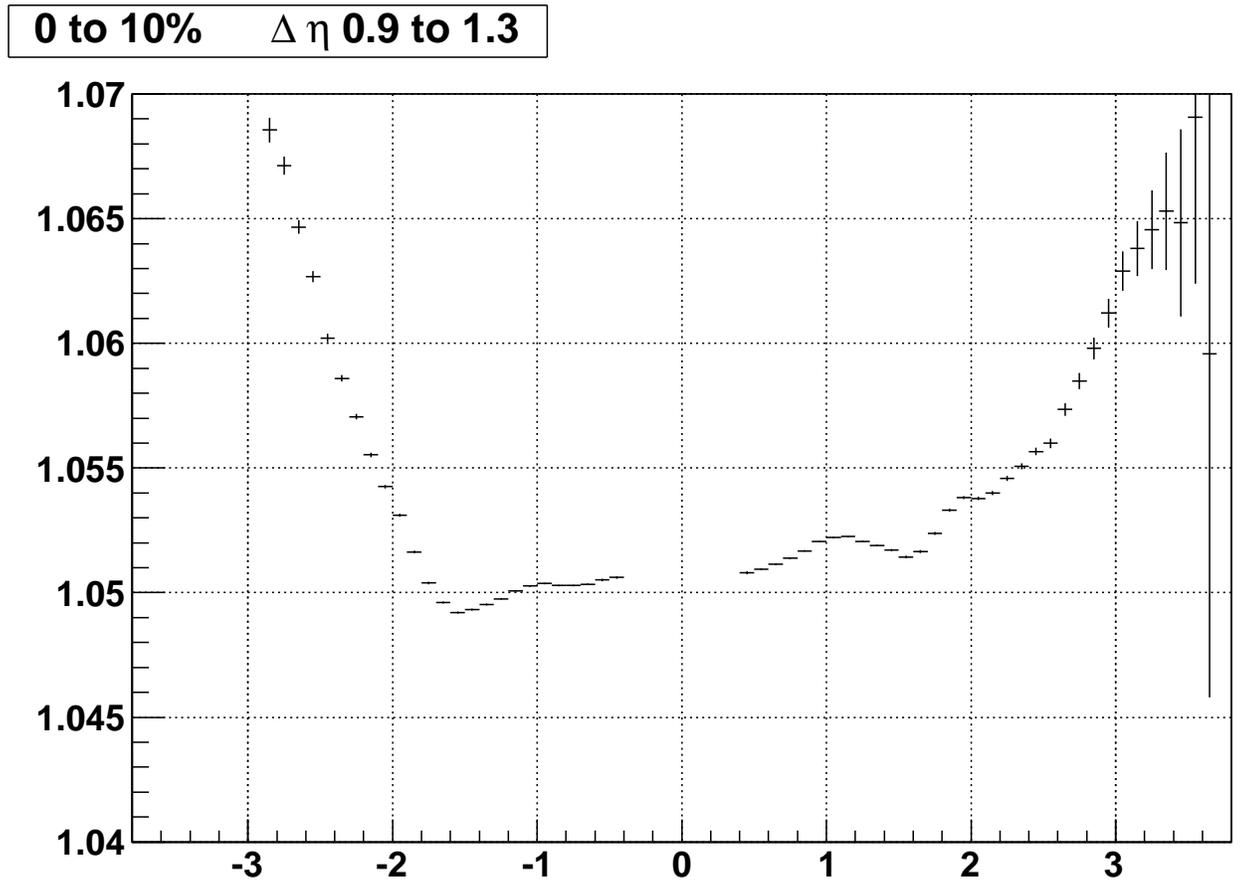}}
\end{center}
\caption{ Correlation function of pairs formed to exhibit the 
effects of longitudinal momentum boosts by the color electric field as defined 
in the text. For this correlation we are using real 0-10\% central Au + Au 
collisions at $\sqrt{s_{NN}} =$ 200 GeV data. The aligned pairs positive values
are larger by $\sim$0.2\% between 1.0 $<$ $\Delta \eta_1 + \Delta \eta_2$ $<$ 
2.0 than between -2.0 $<$ $\Delta \eta_1 + \Delta \eta_2$ $<$ -1.0). 
This means there are more pairs of pairs aligned in the same direction 
compared to pairs not aligned as predicted by the color electric field.}
\label{fig17}
\end{figure}

\begin{figure}
\begin{center}
\mbox{
   \epsfysize 7.0in
   \epsfbox{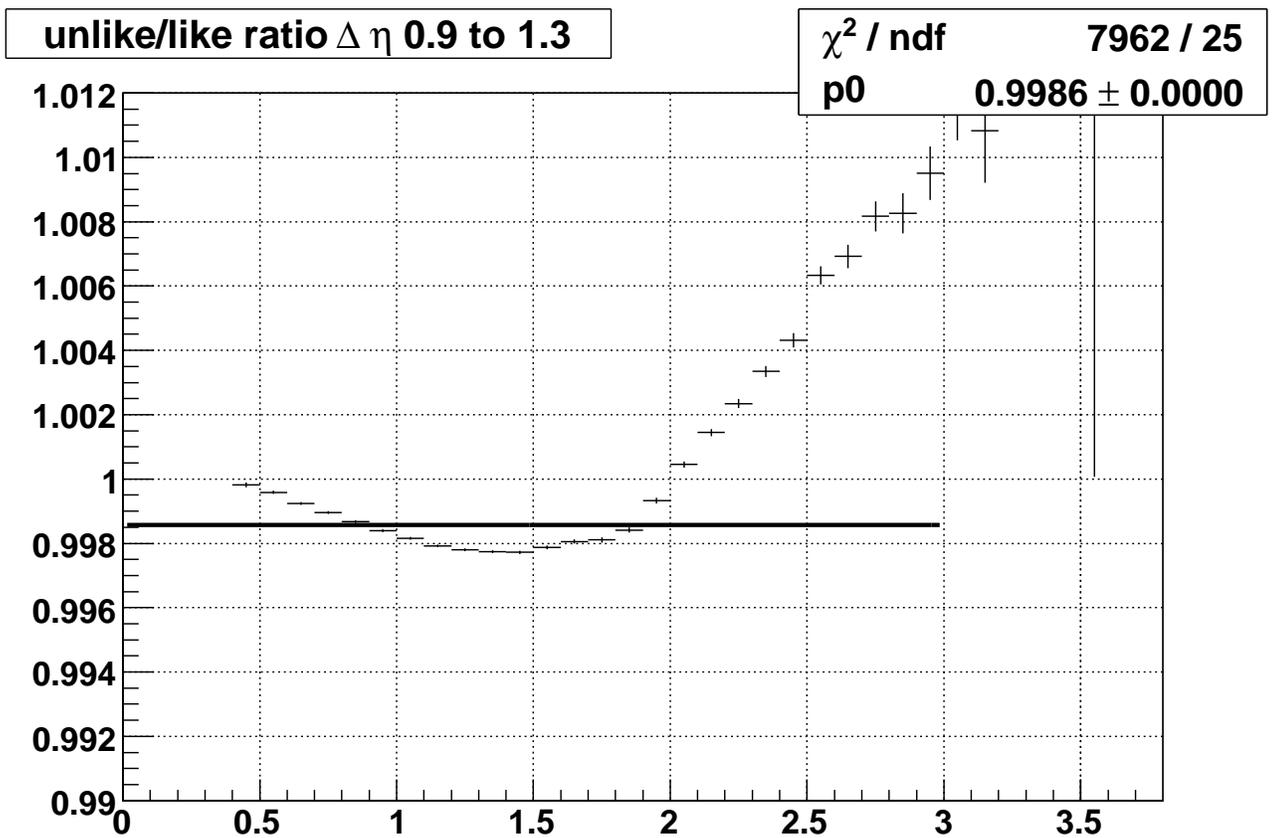}}
\end{center}
\caption{We can obtain a measure of how significant this difference is by 
forming a ratio of not aligned(unlike) over aligned(like). If there was 
no effect the ratio would be a straight line at 1.0. We see that there is a 
dip of 0.2\% at a $\Delta \eta_1 + \Delta \eta_2$ = 1.4 with an average value
shift of 0.14\% over the expected range.}
\label{fig18}
\end{figure}

\section{Color magnetic pionic measure and data.}
  
After considering the color electric effect we turn to the color magnetic
effect which causes up quarks to rotate around the flux tube axis in one 
direction, while the anti-up quarks rotate in the other direction. So when 
a pair of quarks and anti-quarks are formed they will pick up or lose transverse
momentum. These changes in $p_t$ will be transmitted to the $\pi^+$ and 
$\pi^-$ pairs. In order to observe these differential $p_t$ changes one must
select pairs on one side of the tube in $\phi$ and compare to other pairs
on the other side of the same tube which would lie around $40^\circ$ to 
$48^\circ$ away in $\phi$. We defined a pair as a plus particle and minus 
particle with an opening angle ($\theta$) of $16^\circ$ or less. We are 
also interested in pairs that are directly on the other side of the 
tube. We require they are near in pseudorapidity ($\Delta \eta$ $<$ 0.2). 
The above requirements constrain the four charged particles comprising both
pairs to be contained in the same tube and be close to being directly
on opposite sides of the tube (a final state expanded flux tube).
The difference in $p_t$ changes due to and predicted by the color magnetic 
effect should give the $\pi^+$ on one side of the tube an increased 
$p_t$ and a decreased $p_t$ on the other side, while for the 
$\pi^-$ it will be the other way around. This will lead to an 
anti-alignment between pairs. In Figure 19 we show two pairs which 
would fall into the above cuts. Both pairs are at the limit 
of the opening angle cut $\theta_1$ and $\theta_2$ equal $16^\circ$. 
The $p_t$ of the plus particle for pair number 1 is 1.14 GeV/c, 
while the minus particle is 1.39 GeV/c. Thus $\Delta P_{t1}$ 
is equal to -0.25 GeV/c. The $p_t$ of the plus particle for 
pair number 2 is 1.31 GeV/c, the minus particle is 0.91 GeV/c 
and $\Delta P_{t2}$ is equal to 0.40 GeV/c. Finally the 
mean value shown on Figure 19 is the mid-point between the $\pi^+$
and the $\pi^-$ where one really uses the vector sum of the $\pi^+$
and the $\pi^-$ which moves this point toward the harder pion.
 
Considering the above cuts we defined a correlation function where we combine 
pairs each having a $\Delta P_t$. Our variable is related to the sum of the 
absolute values of the individual $\Delta P_t$'s 
($\vert \Delta P_{t1} \vert +\vert \Delta P_{t2} \vert$). We assign a sign
to this sum such that if the sign of the individual $\Delta P_t$'s are
the same it is a plus sign, while if they are different it is a minus sign.
For the flux tube the color magnetic field extends over a large pseudorapidity
range where quarks and anti-quarks rotate around the flux tube axis, therefore 
we want to sample pairs at the different sides of the tube making a separation
in $\phi$ ($\Delta \phi$) between $40^\circ$ to $48^\circ$. We are interested
in sampling the pairs on the other side so we require the separation in $\eta$
($\Delta \eta$) be 0.2 or less. For the numerator of the
correlation function we consider all combination of unique pairs
(sign ($\vert \Delta P_{t1} \vert +\vert \Delta P_{t2} \vert$)) from a given 
central Au + Au event divided by a mixed event denominator created from pairs
in different events. We determine the rescale of the mixed event denominator by
considering the number of pairs of pairs for the case $\vert \Delta \eta \vert$
lying between 1.2 and 1.5 plus any value of $\vert \Delta \phi \vert$ for 
events and mixed events so that the overall ratio of this sample numerator to 
denominator is 1. By picking this $\Delta \eta$ bin for all
$\vert \Delta \phi \vert$ we have around the same pair count as the signal
cut with the $\Delta \phi$ correlation of the tubes being washed out.
For a simpler notation let 
(sign ($\vert \Delta P_{t1} \vert +\vert \Delta P_{t2} \vert$)) =
$\Delta P_{t1} + \Delta P_{t2}$ which we plot in the range from -4 to +4 since 
we have an overall $p_t$ range 0.8 to 4.0 GeV/c. Thus the maximum magnitude 
of $\Delta P_t$'s is 3.2 GeV/c which makes  $\Delta P_{t1} + \Delta P_{t2}$ 
have a range of $\pm$6.4. However the larger values near these range limits
occur very rarely.

\begin{figure}
\begin{center}
\mbox{
   \epsfysize 5.0in
   \epsfbox{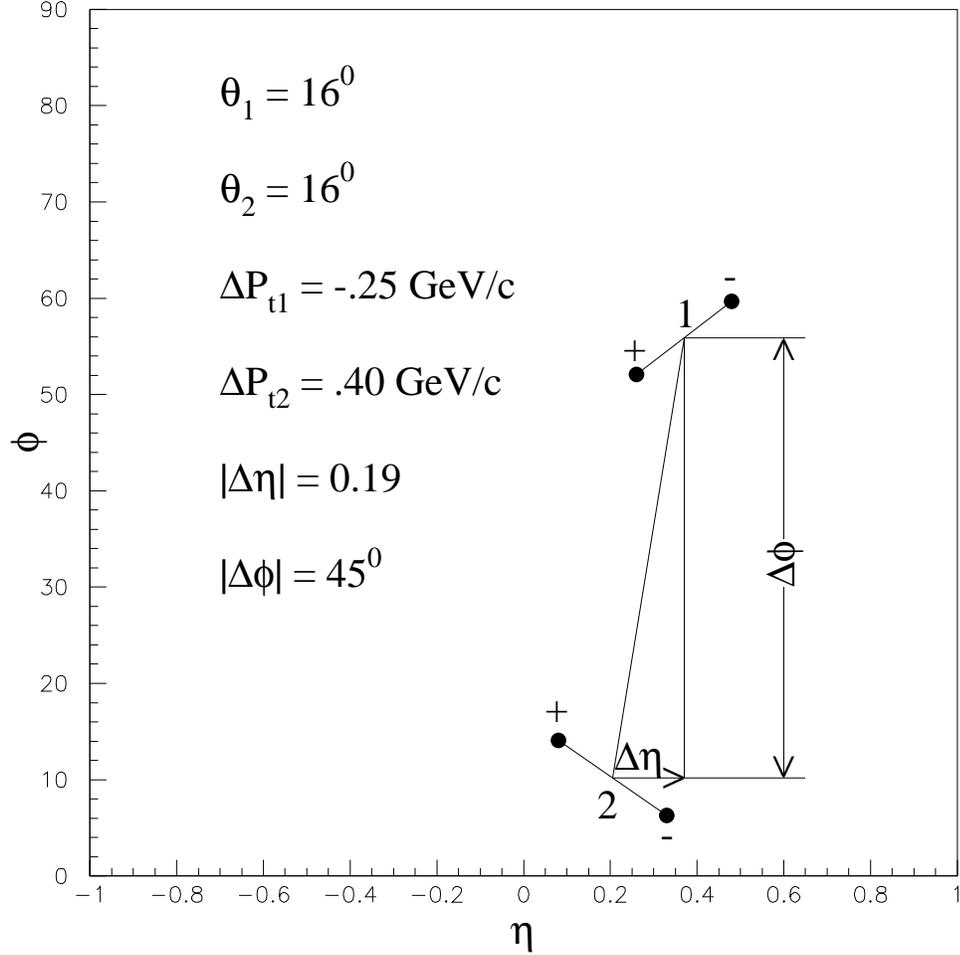}}
\end{center}
\caption{Pairs of pairs selected for forming a correlation 
exhibiting the effect of changes in $p_t$ due to the color magnetic field. The 
largest opening angle $\theta$ for plus minus pairs is $16^\circ$ or less. This
opening angle assures that each pair has a high probability that it arises
from a  quark anti-quark pair. In this figure we have picked two pairs
at this limit ($\theta_1$ = $16^\circ$ and $\theta_2$ = $16^\circ$).
The mid-point for pair 1 and 2 represents the vector sum of pair 1 and 2
which moves toward the harder particle when the momenta differ. These
mid-points are chosen to have $40^\circ$ $<$ $\vert \Delta \phi \vert$ $<$
$48^\circ$ in order for the pairs to be on opposite sides of the tube. 
Since we are interested in pairs directly across the tube we make the 
$\Delta \eta$ separation be no more than 0.2. The difference in $p_t$
for pair 1 is $\Delta P_{t1}$ = -0.25 GeV/c, while the difference in $p_t$
for pair 2 is $\Delta P_{t2}$ = 0.40 GeV/c. The minus sign for 1 follows
from the fact that the plus particle has 1.14 GeV/c and the minus particle
has 1.39 GeV/c. The plus sign for 2 follows from the fact that the plus 
particle has 1.31 GeV/c and the minus particle has 0.91 GeV/c.}
\label{fig19}
\end{figure}

\clearpage

In Figure 20 we show the correlation function of opposite sign
charged-particle-pairs paired and binned by the variable 
$\Delta P_{t1} + \Delta P_{t2}$ with a cut $\vert \Delta \eta \vert$ 
less than 0.2 between the vector sums
of the two pairs, and with $40^\circ$ $<$ $\vert \Delta \phi \vert$ $<$ 
$48^\circ$. The events are generated by the PBM\cite{PBM} and are charged 
particles of 0.8 $<$ $p_t$ $<$ 4.0 GeV/c, with $\vert \eta \vert$ $ <$ 1, from 
Au + Au central (0-10\%) collisions at $\sqrt{s_{NN}} =$ 200 GeV. Since we 
select pairs of pairs which are near each other in $\phi$ 
($40^\circ$ to $48^\circ$) they all together pick up the tube correlation and 
thus these pairs of pairs overall show about a 0.4\% correlation. In the -4.0 
$<$ $\Delta P_{t1} + \Delta P_{t2}$ $<$ -1.0 region the correlation increases 
from 0.4\% to 1\%, while in the 1.0 $<$ $\Delta P_{t1} + \Delta P_{t2}$ 
$<$ 4.0 region the correlation decreases from 0.4\% to -0.2\%. This means the 
pairs of pairs are anti-aligned at a higher rate than aligned. This 
anti-alignment is what is predicted by the color magnetic field effect 
presented above. In fact the anti-alignment increases with $p_t$ as the ratio 
tube particle to background increases.

We apply the magnetic analysis to 0-10\% central Au + Au collisions at 
$\sqrt{s_{NN}} =$ 200 GeV. The error bars on the correlation is such that
we need increase our $\Delta\phi$ cut to 
$32^\circ$ $<$ $\vert \Delta \phi \vert$ $<$$48^\circ$. The STAR TPC efficiency
varied with azimuthal angle because of sector boundaries in the TPC. 
12 boundaries generate effects every $30^\circ$. Pairs of pairs have a lower
efficiency if separated by $30^\circ$.  Pairs that lie with a $\phi$ separation
between $25^\circ$ and $35^\circ$ have both pairs going into sector boundaries 
in twelve orientations in the TPC. This efficiency is less than other angles 
where one of its pairs will go into sector boundaries at 24 orientations in the 
TPC. Thus the numerator of the correlation is reduced more in the TPC data.
In figure 21 we show that aligned pairs positive $\Delta P_{t1} + 
\Delta P_{t2}$ is 0.1\% less than not aligned pairs negative $\Delta P_{t1} 
+ \Delta P_{t2}$. This is about half the value that we had simulated, and 
shows that this alignment effect is present in the data. The average value
of the correlation has been reduced because of the TPC efficiency. We can 
obtain a measure of how significant this difference is by forming a ratio of 
not aligned(unlike) over aligned(like)(see Figure 22). If there was no effect 
the ratio would be a straight line at 1.0. We see that there is a rise of 
0.1\% in a parabolic shape which is a good fit at a $\Delta P_{t1} + 
\Delta P_{t2}$  = 2.0.

\begin{figure}
\begin{center}
\mbox{
   \epsfysize 6.4in
   \epsfbox{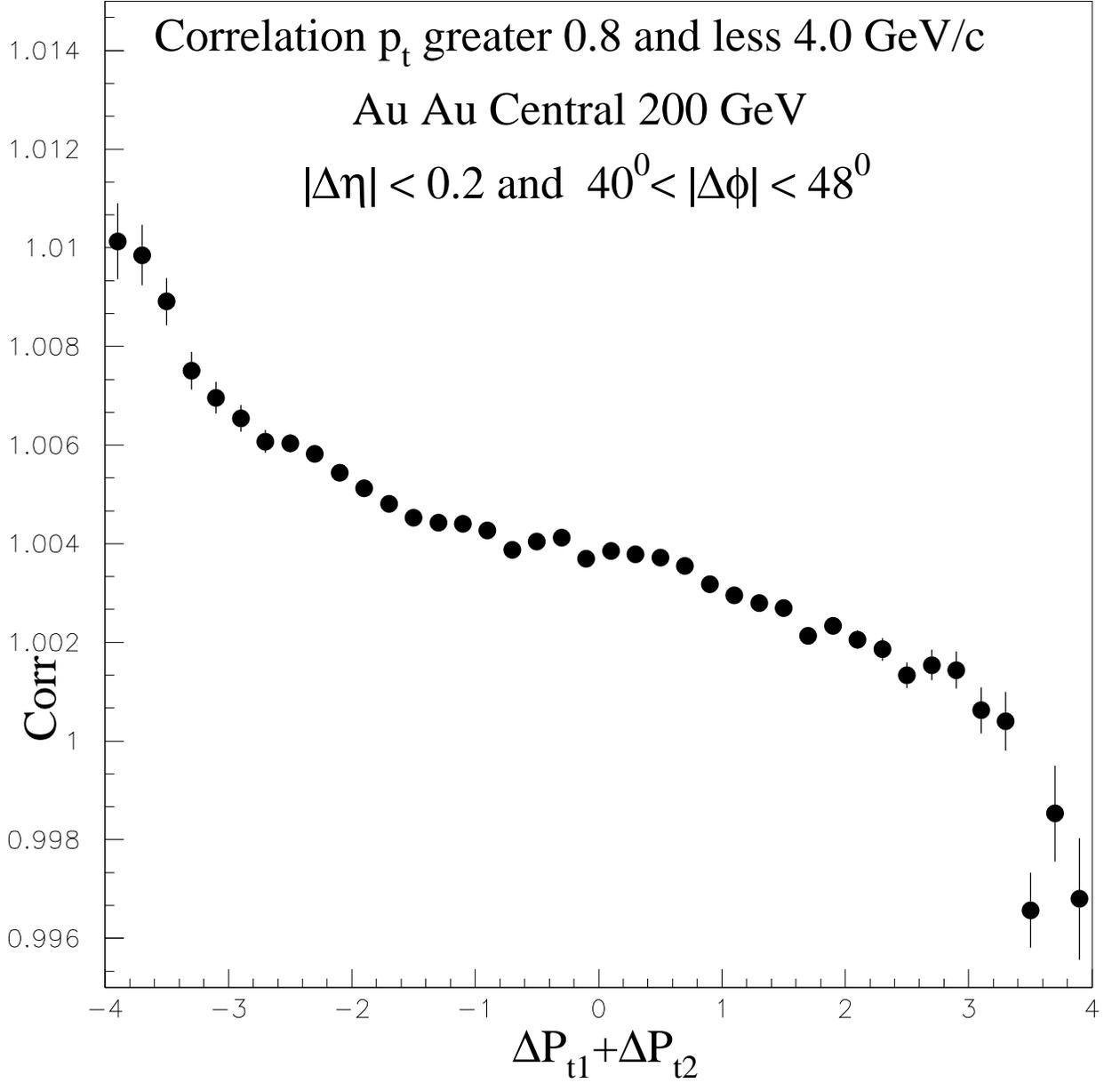}}
\end{center}
\caption{ Correlation function of pairs of pairs formed for 
exhibiting the effect of changes in $p_t$ due to the color magnetic field as 
defined in the text. This correlation shows that there are more anti-aligned 
pairs. The correlation is larger by $\sim$1.2\%  for 
$\Delta P_{t1} + \Delta P_{t2}$ = -4.0 compared to 
$\Delta P_{t1} + \Delta P_{t2}$ = 4.0. 
$\Delta P_{t1} + \Delta P_{t2}$ is equal to 
$\vert \Delta P_{t1} \vert +\vert \Delta P_{t2} \vert$ 
where the sign is also explained in the text.}
\label{fig20}
\end{figure}

\begin{figure}
\begin{center}
\mbox{
   \epsfysize 6.4in
   \epsfbox{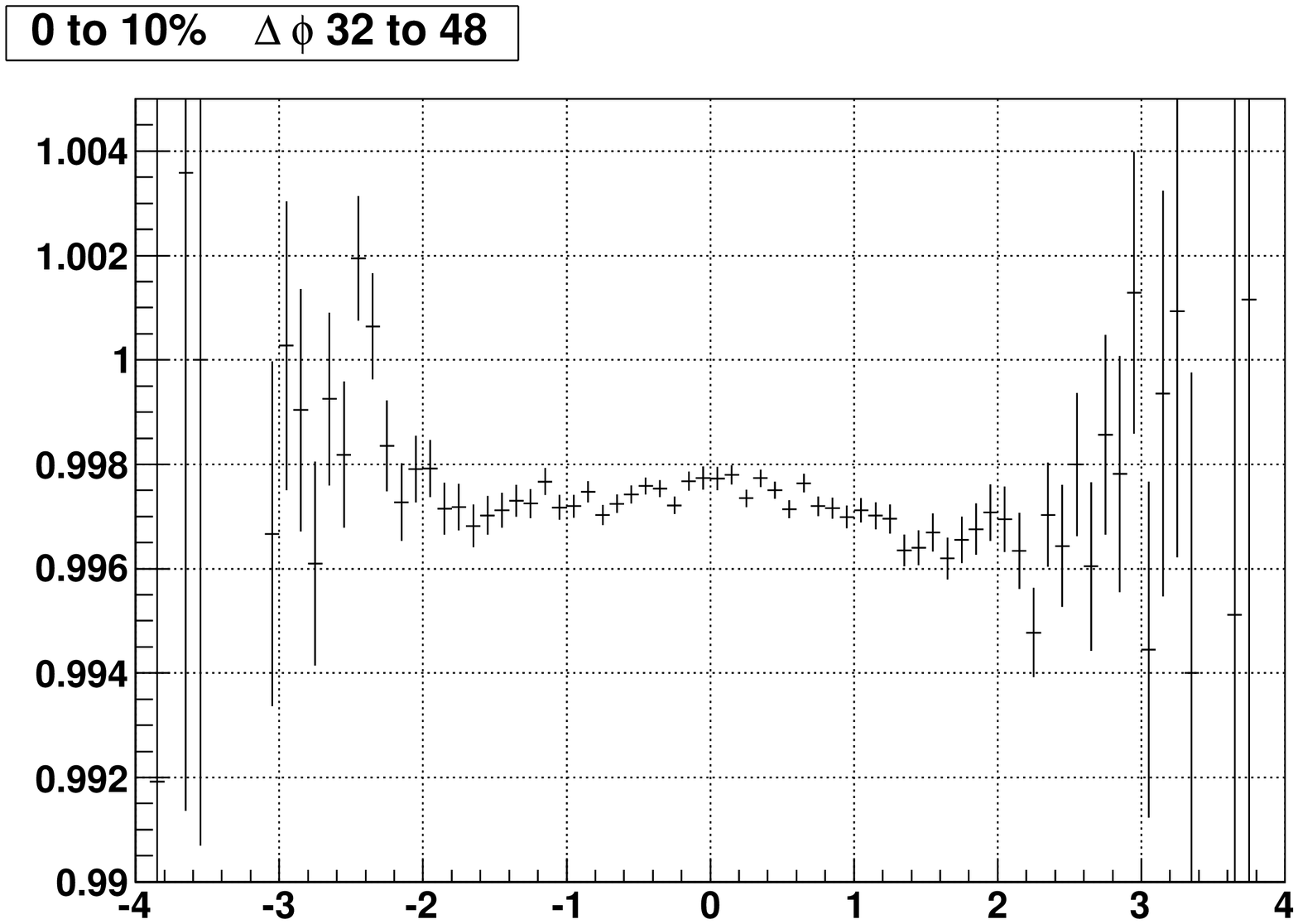}}
\end{center}
\caption{ Correlation function of pairs of pairs formed for 
exhibiting the effect of changes in $p_t$ due to the color magnetic field as 
defined in the text. This correlation shows that there are more anti-aligned 
For this correlation we are using real 0-10\% central Au + Au 
collisions at $\sqrt{s_{NN}} =$ 200 GeV data. The not aligned pairs negative
values are larger by $\sim$0.1\% at -2.0 $\Delta P_{t1} + \Delta P_{t2}$ .
This means there are more pairs of pairs aligned in the opposite direction 
compared to pairs aligned as predicted by the color magnetic field.}
\label{fig21}
\end{figure}

\begin{figure}
\begin{center}
\mbox{
   \epsfysize 6.4in
   \epsfbox{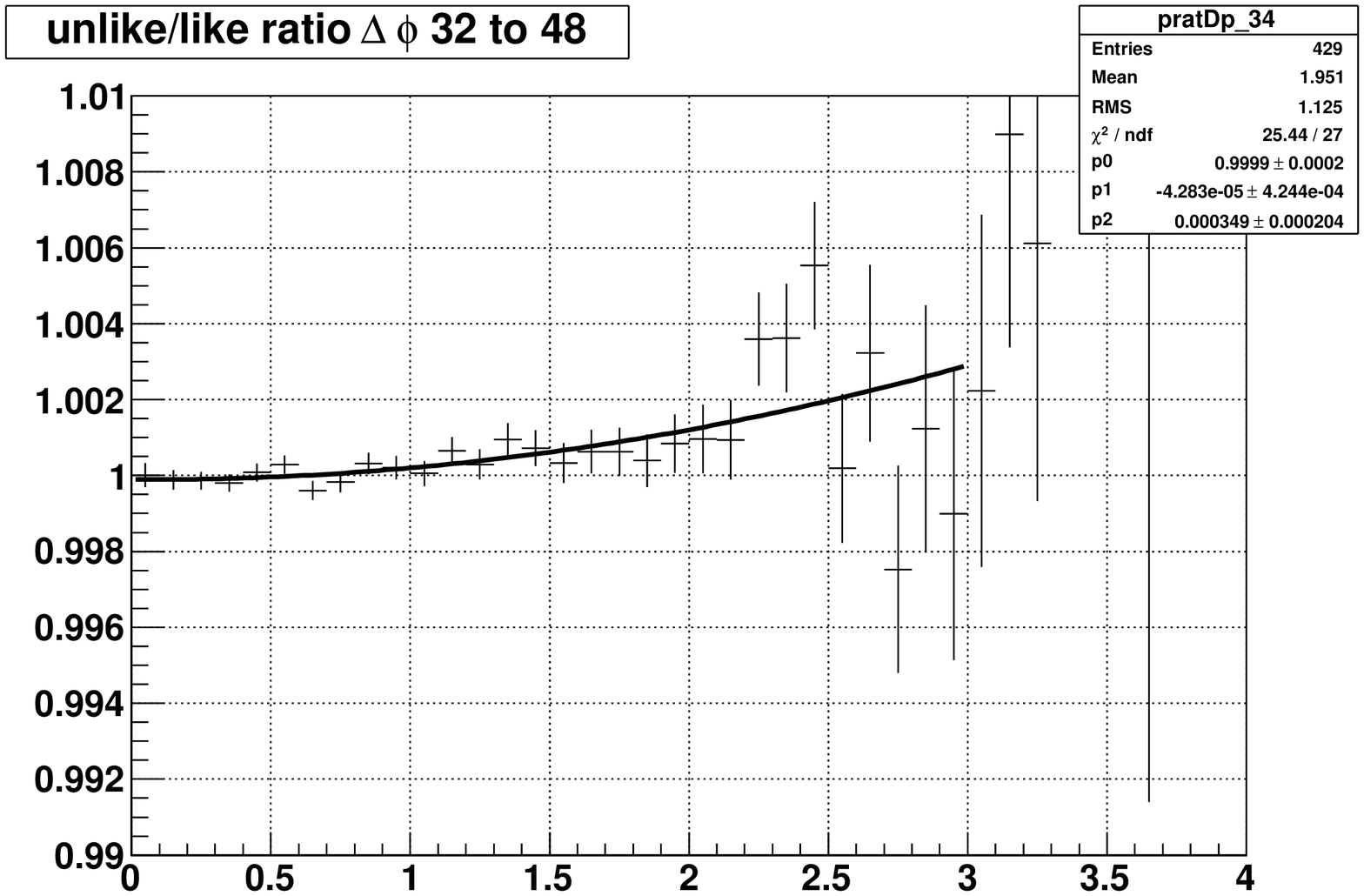}}
\end{center}
\caption{We can obtain a measure of how significant this difference is by 
forming a ratio of not aligned(unlike) over aligned(like). If there was 
no effect the ratio would be a straight line at 1.0. We see that there is a 
rise of 0.1\% in a parabolic shape which is a good fit at a 
$\Delta P_{t1} + \Delta P_{t2}$  = 2.0.}
\end{figure}

\section{Summary and Discussion}

In this paper we discuss a Glasma Flux Tube Model(GFTM)\cite{Dumitru}
which has tubes of parallel strong color QCD electric and magnetic fields that 
are generated by the initial conditions of Au + Au central collisions at 
$\sqrt{s_{NN}}$=200 GeV.  We summarize the assumptions made and the reasoning 
that led to the development and construction of GFTM, which successfully 
explained the charge-particle-pair correlations in the central (0-10\% 
centrality) $\sqrt{s_{NN}} =$ 200 GeV Au Au data\cite{centralproduction}. 

In the GFTM a flux tube is formed right after the initial collision 
of the Au Au system. This flux tube extends over many units of 
pseudorapidity ($\Delta \eta$). A blast wave gives the tubes near the surface 
transverse flow. Initially the transverse space is filled with flux 
tubes of large longitudinal extent but small transverse size $\sim$$Q^{-1}_s$. 
The flux tubes that are near the surface of the fireball get the largest 
radial flow and are emitted from the surface. These partons shower and the 
higher $p_t$ particles escape the surface and do not interact. $Q_s$ is around 
1 GeV/c thus the size of the flux tube is about 1/4 fm initially. 
The flux tubes near the surface are initially at a radius $\sim$5 fm. The 
$\phi$ angle wedge of the flux tube $\sim$ 1/20 radians or $\sim$$3^\circ$.
The flux tubes on the surface turn out to be on the average 12 in number. They
form an approximate ring about the center of the collision see Figure 1. The 
twelve tube ring creates the average behavior of tube survival near the 
surface of the expanding fire ball of the blast wave. The final state surface 
tubes that emit the final state particles at kinetic freezeout for this paper
are given by the PBM\cite{PBM}. One should note that the blast wave surface is 
moving at its maximum velocity at freezeout (3c/4). 

We explored the charge dependence correlation where we separate pairs into 
like sign pairs and unlike sign pairs. In Figure 11 we show the
unlike sign pair correlation of the single flux tube and the background 
subtracted data. In Figure 12 we show the same for like sign pairs.
We see in both Figure 11 and Figure 12, that the correlation from each pair
choice is well described by GFTM. We see at small $\Delta \phi,\Delta \eta$
there is a local bump for unlike sign pairs, while there is negative bump or
dip for like sign pairs. This dip or hole torn is caused by QCD field shown
in Figure 13. This field focus unlike charge pairs into the same 
$\Delta \phi,\Delta \eta$ and defocus like charge pairs out of this same
region.

Strong CP violation (Chern-Simons topological charge) is
treated in Sec. 6.

We develop a predictive pionic measure of the strong CP Violation. 
The GFTM flux tubes are made up of longitudinal color 
electric and magnetic fields which generate topological Chern-Simons 
charge\cite{Simons} through the $F \widetilde F$ term that becomes a source of 
CP violation. The color electric field which points along the flux tube axis 
causes an up quark to be accelerated in one direction along the beam axis, 
while the anti-up quark is accelerated in the other direction. So when a pair 
of quarks and anti-quarks are formed they separate along the beam axis leading 
to a separated $\pi^+$ and $\pi^-$ pair along this axis. The color 
magnetic field which also points along the flux tube axis (which is parallel 
to the beam axis) causes an up quark to rotate around the flux tube axis 
in one direction, while the anti-up quark rotates in the other direction. 
So when a pair of quarks and anti-quarks are formed they will pick up or lose 
transverse momentum. These changes in $p_t$ will be transmitted to the 
$\pi^+$ and $\pi^-$ pairs.

The above pionic measures of strong CP violation are used to form correlation
functions based on four particles composed of two pairs which are opposite 
sign charge-particle-pairs that are paired and binned. These four particle 
correlations accumulate from tube to tube by particles that are
pushed or pulled (by the color electric field) and rotated (by the color 
magnetic field) in a right or left handed direction. The longitudinal 
color electric field predicts aligned pairs in a pseudorapidity or 
$\Delta \eta$ measure. The longitudinal color magnetic field predicts 
anti-aligned pairs in a transverse momentum or $\Delta P_t$ measure. 
We observe in the STAR TPC $\sqrt{s_{NN}}$ = 200 GeV central Au Au 
collision data that these correlations are as expected and is a strong
confirmation of this theory(see Sec. 7 and Sec. 8). The much larger 
unlike-sign pairs than like-sign pairs in the data and the strong dip of the 
CI correlation at small $\Delta \eta$ (see Sec. 5) shows very strong
evidence supporting the color electric alignment prediction
in the $\sqrt{s_{NN}}$ = 200 GeV central Au Au collision data analyses at
RHIC\cite{PBM,centralproduction}. This highly significant dip 
($\sim$$14\sigma$) means that like-sign pairs are removed as one approaches 
the region $\Delta \eta$ = $\Delta \phi$ = 0.0. Thus this is the strongest
evidence for the predicted effect of the color electric field.

\section{Acknowledgments}

This research was supported by the U.S. Department of Energy under Contract No.
DE-AC02-98CH10886. The author thanks Sam Lindenbaum and William Love for 
valuable discussion and Bill for assistance in production of figures. It is 
sad that both are now gone.


\begin{thebibliography}{99}
\bibitem{Dumitru} A.~Dumitru, F.~Gelis, L.~McLerran and R.~Venugopalan, 
Nucl. Phys. A 810 (2008) 91.
\bibitem{CGC} L.~McLerran and R.~Venugopalan, Phys. Rev. D 49 (1994) 2233;  
Phys. Rev. D 49 (1994) 3352;  Phys. Rev. D 50 (1994) 2225.
\bibitem{Lappi} T.~Lappi and L.~McLerran, Nucl. Phys. A 772 (2006) 200.
\bibitem{Gelis} F.~Gelis and R.~Venugopalan, Acta Phys. Polon. B 37 (2006)
 3253.
\bibitem{Simons} D.~Kharzeev, A.~Krasnitz and R.~Venugopalan, Phys. Lett. B
545 (2002) 298.
\bibitem{Romatschke1} P.~Romatschke and R.~Venugopalan, Phys. Rev. D 74 (2006) 045011.
\bibitem{Gavin} S.~Gavin, L.~McLerran and G.~Moschelli, Phys. Rev. C 79 (2009) 051902. 
\bibitem{PBM} S.J.~Lindenbaum, R.S.~Longacre, Eur. Phys. J. C. 49 (2007)767.
\bibitem{PBMtoGFTM} S.J.~Lindenbaum, R.S.~Longacre, arXiv:0809.2286[Nucl-th].
\bibitem{pythia} T.~Sjostrand, M.~van Zijil, Phys. Rev. D 36 (1987) 2019.
\bibitem{HBT} J.~Adams {\it et al.}, Phys. Rev. C 71 (2005) 044906, S.S.~Adler
{\it et al.}, Phys. Rev. Lett. 93 (2004) 152302.
\bibitem{centralproduction} J.~Adams {\it et al.}, Phys. Rev. C 75, 034901 (2007).
\bibitem{PBME} S.J.~Lindenbaum and R.S.~Longacre, Phys. Rev. C 78 (2008) 054904.
\bibitem{centralitydependence} B.I.~Abelev {\it et al.}, arXiv:0806.0513[nucl-ex]. 
\bibitem{Alver} B.~Alver and G.~Roland, Phys. Rev. C 81 (2010) 054905.
\bibitem{Witten} C.~Vafa, E.~Witten, Phys. Rev. Lett. 53 (1984) 535, Nucl. 
Phys. B 234 (1984) 173.
\bibitem{Tytgat} D.~Kharzeev, R.D.~Pisarski, M.H.G.~Tytgat, Phys. Rev. Lett. 81
(1998) 512.
\bibitem{Pisarski} D.~Kharzeev, R.D.~Pisarski, Phys. Rev. D 61 (2000) 111901.
\end{thebibliography}
\end{document}